\documentclass{aa} 

\usepackage[varg]{txfonts}
\usepackage{natbib}
\bibpunct{(}{)}{;}{a}{}{,} 
\usepackage{graphicx}
\usepackage{hyperref}
\usepackage{color}
\usepackage{pgfkeys}
\newcommand{\setvalue}[1]{\pgfkeys{/variables/#1}}
\newcommand{\getvalue}[1]{\pgfkeysvalueof{/variables/#1}}
\newcommand{\declare}[1]{%
 \pgfkeys{
  /variables/#1.is family,
  /variables/#1.unknown/.style = {\pgfkeyscurrentpath/\pgfkeyscurrentname/.initial = ##1}
 }%
}
\declare{}
\newcommand\figcomp[4]{
\begin{figure*}
\centering
\includegraphics[width=0.72\hsize,clip]{Figures/CO_images/composite/#1/ANALYSIS/#2/#1-#2-composite.pdf}
\caption{Observations of #3 emission in the #4 disk.
Shown are the channel map of the observed line brightness distribution (left), the moment-zero map (top right), and the integrated spectrum (bottom right). The channel map shows 16 velocity channels with a step of $0.5$~km\,s$^{-1}$ in the [-4.5, +4.5]~km\,s$^{-1}$ range around the systemic velocity. The contour lines start at $2$~K, with a step of $2$~K. The color bar shows line brightness temperatures (K). The contour lines in the moment-zero plot start at 3, 6, and $9\sigma$, with a step of $6\sigma$ afterward. The $1\sigma$ rms noise (mJy\,km\,s$^{-1}$) is shown in the upper left corner of the moment-zero plot, and the synthesized beam is depicted by the dark ellipse in the left bottom corner.}
\label{fig:#1-#2_comp}
\end{figure*}
}
\newcommand\figkepl[4]{
\begin{figure*}
\sidecaption
  \includegraphics[width=12cm,clip]{Figures/CO_images/kepler/#1/ANALYSIS/#2/#1-#2-kepler.pdf}
\caption{Observations of #3 emission in the #4 disk. Shown is a pixel-deprojected Keplerian plot consisting of the three panels: (left) the radial profile of the line brightness temperature (K), with observations and errorbars in black, and profile derived from the best-fit disk model in red, (top) the integrated spectrum (black line) overlaid with the best-fit Gaussian profile (red line), and (bottom right) aligned and stacked line intensity (K) as a function of disk radius (Y-axis; au) and velocity (X-axis; km\,s$^{-1}$). The color bar units are in Kelvin.}
\label{fig:#1-#2_kepl}
\end{figure*}
}

\newcommand{\textsg}[1]{\textcolor{black}{#1}}
\newcommand{\textds}[1]{\textcolor{black}{#1}}
\newcommand{\upd}[1]{\textcolor{black}{#1}}
\newcommand{\sgupd}[1]{\textcolor{black}{#1}}

\begin{document} 

  \title{PRODIGE - Planet-forming disks in Taurus with NOEMA}
  \subtitle{I. Overview and first results for $^{12}$CO, $^{13}$CO, and C$^{18}$O}
  \author{D. Semenov\inst{1,2}
  \and Th. Henning\inst{1}
  \and S. Guilloteau\inst{3,4}
  \and G. Smirnov-Pinchukov\inst{1}
  \and A. Dutrey\inst{3,4}
  \and E. Chapillon\inst{5}
  \and V. Pi\'etu\inst{5}
  \and R. Franceschi\inst{1}
  \and K. Schwarz\inst{1}
  \and S. van Terwisga \inst{1} 
  \and L. Bouscasse \inst{5}
  \and P. Caselli \inst{6}
  \and C. Ceccarelli \inst{7}
  \and N. Cunningham \inst{7}
  \and A. Fuente \inst{8}
  \and \textds{C. Gieser \inst{6}}
  \and T.-H. Hsieh \inst{6}
  \and A. Lopez-Sepulcre \inst{5,7}
  \and D. M. Segura-Cox \inst{9,6}\thanks{NSF Astronomy and Astrophysics Postdoctoral Fellow}
  \and J. E. Pineda \inst{6}
  \and M. J. Maureira  \inst{6}
  \and Th. M\"{o}ller \inst{10}
  \and M. Tafalla \inst{8}
  \and M. T. Valdivia-Mena \inst{6}
  }

\institute{Max-Planck-Institut f\"{u}r Astronomie (MPIA), K\"{o}nigstuhl 17, D-69117 Heidelberg, Germany\\
\email{semenov@mpia.de}
\and Department of Chemistry, Ludwig-Maximilians-Universit\"{a}t, Butenandtstr. 5-13, D-81377 M\"{u}nchen, Germany
\and LAB, Universit\'e de Bordeaux, B18N, All\'ee Geoffroy, Saint-Hilaire, CS 50023, 33615 Pessac Cedex
\and CNRS, Universit\'e de Bordeaux, B18N, All\'ee Geoffroy, Saint-Hilaire, CS 50023, 33615 Pessac Cedex
\and IRAM, 300 Rue de la Piscine, F-38046 Saint Martin d'H\`{e}res, France
\and Max-Planck-Institut f\"{u}r extraterrestrische Physik (MPE), Gie{\ss}enbachstr. 1, D-85741 Garching bei M\"{u}nchen, Germany
\and IPAG, Universit\'{e} Grenoble Alpes, CNRS, F-38000 Grenoble, France
\and Centro de Astrobiología (CAB), CSIC-INTA, Ctra Ajalvir Km 4, Torrejón de Ardoz, 28850 Madrid, Spain
\and Department of Astronomy, The University of Texas at Austin, 2500 Speedway, Austin, TX, 78712, USA
\and I. Physikalisches Institut, Universität zu K\"{o}ln, Z\"{u}lpicher Str. 77, 50937 K\"{o}ln, Germany
}

\titlerunning{PRODIGE - Protoplanetary disks in Taurus with NOEMA. I}
\authorrunning{D.\ Semenov\ et al.}

   \date{\today}

\abstract
{The physics and chemistry of planet-forming disks are far from being fully understood. \upd{To make further progress, both broad line surveys and observations of individual tracers in} a statistically significant number of disks are required.}
{Our aim is to perform a line survey of \upd{eight planet-forming Class~II disks in Taurus} with the IRAM NOrthern Extended Millimeter Array (NOEMA), as a part of the MPG-IRAM Observatory Program PRODIGE (PROtostars and DIsks: Global Evolution; PIs: P.~Caselli and Th.~Henning). 
}
{Compact and extended disks around T~Tauri stars CI, CY, DG, DL, DM, DN, IQ~Tau, and UZ~Tau~E are observed \upd{in $\sim 80$ lines from $>20$ C-, O,- N-, and S-bearing species.} The observations \upd{in four spectral settings at 210-280 GHz with a $1\sigma$ rms sensitivity} of $\sim 8-12$~mJy/beam at a $0.9\arcsec$ and 0.3~km\,s$^{-1}$ resolution \upd{will be completed in 2024. 
The $uv$ visibilities} are fitted with the DiskFit model to obtain key stellar and disk properties.}
{\upd{In this first paper, the combined $^{12}$CO, $^{13}$CO, and C$^{18}$O~$J=2-1$ data are presented. We find that the CO fluxes and disk masses inferred from dust continuum tentatively correlate with the CO emission sizes. We constrained dynamical stellar masses, geometries, temperatures, the CO column densities, and gas masses for each disk.} 
The best-fit temperatures \upd{at 100~au} are $\sim 17-37$~K, \upd{and decrease radially with the power-law exponent $q \sim 0.05-0.76$. The inferred CO column densities decrease radially with the power-law exponent $p \sim 0.2-3.1$. The gas masses estimated from $^{13}$CO~(2-1) are $\sim 0.001 - 0.2~M_\odot$.}}
{
\upd{Using NOEMA, we confirm the presence of temperature gradients in our disk sample. The best-fit CO column densities point to severe CO freeze-out in these disks. The DL~Tau disk is an outlier, and has either stronger CO depletion or lower gas mass than the rest of the sample. The CO isotopologue ratios are roughly consistent with the observed values in disks and the low-mass star-forming regions. The high $^{13}$CO/C$^{18}$O ratio of $\sim 23$ in DM~Tau could be indicative of strong selective photodissociation of C$^{18}$O in this disk. }
} 


   \maketitle
%

\section{Introduction}
\label{sec:intro}
A diversity of discovered exoplanets calls for a better understanding of the physical and chemical processes in their natal planet-forming disks \citep[e.g.,][]{Mordasini_ea12,Turrini_ea21,Molliere_ea22}. \upd{Recently, young planets have been imaged}  inside the gas-rich PDS~70 disk for the first time \citep[e.g.,][]{Keppler_ea18a,Haffert_19,Facchini_ea21_PDS70}. However, despite \upd{extensive progress achieved with the powerful Very Large Telescope (VLT), {\em Spitzer}, James Webb Space Telescope (JWST), {\em Herschel}, Atacama Large Millimeter/Submillimeter Array (ALMA), extended Jansky Very Large Array (eVLA), and NOrthern Extended Millimeter Array (NOEMA)}, our understanding of the disk physics and chemical composition is still limited.
\upd{Among the key disk properties, we have only begun to determine their (1) mass measurements, (2) dust-to-gas ratios, (3) temperature structures, and (4) elemental ratios and molecular abundances in the gas and ice phases \citep[e.g.,][]{Andrews_2020,MAPS1_Oeberg21,Benisty_ea_PPVII_2022,Manara_ea_PPVII_2022,Miotello_ea_PPVII_2022,Oeberg_ea23_ARAA}.} 

\upd{The majority of the disk masses have been inferred from the dust continuum data, often by using fixed dust emissivities and/or dust temperatures, and ill-constrained gas-to-dust ratios \citep[see, e.g., ][and references therein]{Bergin_Williams18_masses,Manara_ea_PPVII_2022}}. 
\upd{Gas masses in a limited number of disks have been obtained by more direct methods. By observing HD emission with {\em Herschel} and performing disk temperature modeling,} the gas masses of TW~Hya, \object{DM Tau}, and GM~Aur disks have been \upd{directly} inferred \citep[][]{Bergin_ea13,McClure_ea16,Trapman:2017tn}. \upd{Using CO emission observed with} ALMA, \citet{Lodato_ea22} have constrained the self-gravity of the IM~Lup and GM~Aur disks, and derived gas masses of $\sim 0.1M_{\odot}$ and $\sim 0.26M_{\odot}$, respectively. \citet{Yoshida_ea22} have measured the pressure broadening of the $^{12}$CO~(3-2) line and inferred high gas densities in the inner TW~Hya disk midplane. \upd{Modeling the dust emission sizes at multiple wavelengths has also provided constraints on the gas surface densities, suggesting that the disks could be more massive than previously thought \citep{Birnstiel_ea18,Powell_ea19,Trapman_ea20,Franceschi_ea21,Tazzari_ea21b}.} 

Disk temperatures at distinct vertical layers are usually probed via a combination of the optically thick and thin lines, such as CO, \upd{and they often show the presence of radial and vertical gradients} \citep[e.g.,][]{DDG03,Williams_Best14,Cleeves_ea16,Miotello_ea18a,Teague_ea18b,2021_MAPS_XVII_in_press,Calahan_ea21,MAPS20_Schwarz_ea21,Miotello_ea_PPVII_2022}.
\upd{For edge-on or well-resolved disks, the vertical temperature profile can be derived directly from the line} observations \citep[e.g.,][]{Dutrey_ea17,Pinte_ea18b,2021_MAPS_IV_in_press,PanequeCarreno_ea22a}. Such studies require combined physical, chemical, and radiative transfer modeling, and hence considerable effort to reach a balance between the model feasibility and its computational costs \citep[][]{Haworth_ea16}. 
 
\upd{In addition to temperatures, turbulent properties of the gas in disks can be estimated from the high signal-to-noise ratio (S/N) line data by accounting for the Keplerian shear and thermal broadening, which requires a good understanding of the underlying thermal structures} \citep[e.g.,][]{Guilloteau_ea12a,Teague_ea16,Flaherty_ea20,PanequeCarreno_ea23a}. Subsonic turbulence has only been \upd{firmly detected in the \object{DM Tau} and IM~Lup disks so far}. 
The deviations from Keplerian rotation have been found in various disks, which \upd{could be signatures of embedded planets or} disk winds \citep[e.g.,][]{Teague_ea18a,Pinte_ea19,Boehler_ea21,2021_MAPS_XVIII_in_press,Stadler_ea23}. 

\upd{To better understand the cause of the apparent lack of turbulence in disks, studies of disk ionization are important. Usually, the} combination of the HCO$^+$ and N$_2$H$^+$ isotopologues \upd{are used to probe the disk ionization degree, which is driven by the interstellar or stellar X-rays} and cosmic ray particles \citep[][]{Dutrey_ea07,Oberg_ea11_ionization,Cleeves_ea15,Teague:2015jk,MAPS13_Aikawa_ea21}. \upd{The
impact of the impinging stellar far-ultraviolet (FUV) radiation on disk properties and molecular composition can be characterized via observations of FUV tracers such as} C$_2$H, c-C$_3$H$_2$, CN, HCN, and HNC, combined with disk chemistry models \citep[e.g.,][]{Bergin_ea04,Henning_ea10,Chapillon_ea12,Guzman_ea15,MAPS11_Bergner_ea21}. 

\upd{The inferred abundances of CO isotopologues, light hydrocarbons (e.g., C$_2$H),} and sulfur-bearing species have revealed substantial depletion of the elemental sulfur and oxygen in some disks, and the nonsolar elemental C/O ratios in their molecular layers \citep[e.g.,][]{Bergin:2016ge,Kama_ea16,Semenov_ea18,LeGal_ea19,Miotello_ea19,Fedele_Favre20,RiviereMarichalar_ea20,MAPS7_Bosman_ea21,Cleeves_ea21_tw_hya,MAPS12_LeGal_ea21,Phuong_ea21_ccs,Ruaud_ea21a,Anderson_ea22,Furuya_ea22a,Pascucci_ea23a}. 
A possible reason for the excessive volatile depletion in the disk molecular layers is the chemical processing and sequestration into refractory ices, which sediment toward the disk midplanes along with the growing dust grains  \citep{Favre_ea13_TWHya_CO,Oberg_Bergin16,Krijt_ea20,Eistrup_Henning22,vanClepper_ea_22}.  
\upd{Still, the overall chemical composition and partitioning of the elemental C, O, N, and S into various species remain poorly constrained in disks. Therefore, a homogeneous study of the disk composition in many chemical species is necessary, and to begin with we need the high S/N CO isotopologue data.}

Last but not least, ubiquitous $\sim 1-100$~au substructures, such as inner holes, gaps, rings, \textds{spirals visible} in both the gas and dust emission are indicative of the planet formation, planet-disk interactions, or disk instabilities, which further complicates the analysis and interpretation of the observations \citep[e.g.,][]{HLTau_rings14,Long_ea17,Andrews_ea18a,SmirnovPinchukov_ea20,2021_MAPS_III_in_press,Bae_ea22_PPVII,Jennings_ea22_gaps,Zhang_ea23_gaps}. The ``Molecules with ALMA at Planet-forming Scales'' (MAPS) collaboration has performed the high-resolution, multiband ALMA observations of the IM~Lup, GM~Aur, AS~209, HD~163296, and MWC~480 disks down to the $\sim 10-20$~au scales \citep[][and references therein]{MAPS1_Oeberg21}. The MAPS team has found a variety of small-scale dust and chemical substructures in these disks, characterized their molecular complexity and the C/O ratios, and revealed complex disk dynamics suggestive of the planet-disk interactions and disk winds.

With the goal of better constraining the physical and chemical structures in the Class~0/I and Class~II sources, we have conceived the MPG-IRAM large observing project PRODIGE (PROtostars to DIsks: Global Evolution) at NOEMA. The first studies of the Class~0/I protostars in Perseus have been published in \citet{ValdiviaMena_ea22} and \citet{Hsieh_ea22}. Here, we present the first paper \upd{about Class~II disks in Taurus. This paper focuses on the analysis and modeling of the CO isotopologue data to unveil the geometry, physical structure, and CO depletion of the targeted disks, which are needed for the future analysis and modeling of other emission lines}. The overview of the project, its scientific goals, and sample selection are described in Section~\ref{sec:overview}. The observations and \upd{CO line} data are presented in Sections~\ref{sec:obs} and \ref{sec:res}, respectively. 
The results of the parametric fitting and \upd{empirical trends between various disk and stellar parameters} are presented and discussed in Section~\ref{sec:diss}. Our summary and conclusions are provided thereafter.

\section{PRODIGE overview, scientific goals, and sample}
\label{sec:overview}
The recently upgraded NOEMA interferometer with $12 \times 15$-m antennas operating at $70-280$~GHz with up to $\sim 1\,800$~m long baselines is a powerful facility\footnote{\url{https://www.iram.fr/GENERAL/NOEMA-Phase-A.pdf}} (see Schuster et al.~2024, in prep.). The PolyFiX correlator can be tuned to include up to 64 62.5~kHz spectral windows in dual polarization, or used with a uniform 250~kHz resolution across the full 15.5~GHz bandwidth \citep{gentaz_oliver_2019_3240337}. In the high-resolution mode, one can target four times more molecular lines compared to the ALMA correlator, which makes NOEMA on par with ALMA for the line surveys at $\sim 0.5-1\arcsec$ resolution. 

\upd{In light of these considerations,} we have devised the large program PRODIGE at NOEMA to study in a coherent manner the physics and chemistry at $\sim 100$~au scales in 8 protoplanetary disks in Taurus (PI: Th.~Henning; 260~h) and in 30 low-mass protostars in Perseus (PI: P.~Caselli; 260~h). The observations began in 2020 and will be completed in 2024. Below we describe the science goals, the sample, and the strategy of the Class~II disk part of the PRODIGE project.

\subsection{Sample}
\label{subsec:sample}
\upd{In contrast to the MAPS project that has targeted 2 T~Tauri and 3 Herbig~Ae disks from various star-forming regions (SFRs) \citep{MAPS1_Oeberg21}, we decided to focus on both the large ($>300$~au) and compact ($<300$~au) T~Tauri disks in Taurus SFR.} We chose the T~Tauri systems with the stellar masses of $M_*\lesssim 1M_{\odot}$, outside the $\sim 1.1-2.1M_{\odot}$ mass range targeted in the MAPS project \citep{MAPS1_Oeberg21}. 
By focusing on the disks that have been formed in the same star-forming region from similar initial conditions, we are able to perform a less biased comparison of the disk properties between the sources \citep[][]{Ansdell_ea18a,Luhman_18,Miotello_ea_PPVII_2022}. 

\upd{The Taurus SFR is young, $\sim 1-5$~Myr \citep{Luhman_23}, nearby \citep[$\sim 140-170$~pc;][]{Gaia_DR2}, and has solar-like metallicities \citep{Cartledge_ea06,DOrazi_ea11,Nieva_Przybilla12}, 
which together with a large, $>200$ population of Class~II disks \citep[][]{Esplin_Luhman19} and a lack of massive stars make it an attractive environment to study the evolution of protoplanetary disks.}

To optimize the \upd{available} observing time \upd{at NOEMA}, we decided to target \upd{a limited number of} sources but with \upd{a high enough sensitivity to detect as many weak lines of multiple species as possible.}
Our sample selection criteria were: 1) Class~II sources in Taurus with ages of $\sim 1-5$~Myr, 2) isolated stars without severe foreground extinction ($A_V<3$~mag), 3) T~Tauri stars only, $M_* \sim 0.4-1.0~M_\odot$, 4) compact and \upd{extended disks} without severe substructures visible in the ALMA continuum images \citep{Long_ea18a,Long_ea19_taurus}, 5) disks with 1.3~mm continuum fluxes of $>60$~mJy to allow self-calibration, 6) disks with previous molecular line detections \citep{Guilloteau_ea16}, and 7) at least one disk with measured gas mass via the HD lines.

Thus, we have selected the following eight systems:
\begin{itemize}
    \item \object{DM Tau}: a large, moderately inclined disk ($\sim 35\degr$) with a gas mass measured via the HD emission with {\em Herschel} \citep{Pietu_ea07,McClure_ea16};
    \item \object{DL Tau}, \object{CI Tau}, \object{CY Tau}, \object{IQ Tau}, \object{DN Tau}: less extended disks with inclinations up to $55\degr$ \citep{Guilloteau_ea11a,Long_ea18a};
    \item \object{UZ Tau}~E: a compact, circumbinary disk with a small inner hole and an inclination of $\sim 56\degr$ \citep{Long_ea18a};
    \item \object{DG Tau}: \upd{a Class I/II object with a jet and an outflow, and an inclination of $\sim 32\degr$ \citep[perhaps, part of a wide binary system with DG~Tau~B;][]{Guilloteau_ea11a}.}
\end{itemize}

\upd{Only some of the targeted disks have been observed with ALMA in the CO isotopologue lines with the sensitivity and resolution comparable to our PRODIGE NOEMA survey (in particular, C$^{18}$O). Furthermore, } CY~Tau, DM~Tau, DL~Tau, DN~Tau, and DG~Tau \upd{have been} observed with JWST in Cycle~1 in the framework of the GTO program ``MIRI EC Protoplanetary and Debris Disks Survey'' (GTO 1282; PI: Th.~Henning\footnote{\url{https://www.stsci.edu/jwst/science-execution/program-information.html?id=1282}}). 

The basic properties of our disk sample are summarized in Tables~\ref{tab:stellar_prop} and \ref{tab:disk_prop}. The additional information about individual sources can be found in Appendix~\ref{append:sec:sample}.

\begin{table*}
\caption{Stellar properties.\label{tab:stellar_prop}}
\centering
\begin{tabular}{llllllllll}
\hline\hline
Source       & R.A.      & Dec   & SpTy & $T_{\rm eff}$ & $M_*$ & $\log_{10}(M_\mathrm{acc})$  & $\log_{10}(L_*)$  & Distance  & References\\
             & ($\degr$ $\arcmin$ $\arcsec$) & ($\degr$ $\arcmin$ $\arcsec$) &  & (K)  & ($M_\odot$) & ($\log_{10} [M_\odot$~yr$^{-1}]$) & ($\log_{10} L_\odot$)  & (pc)  & \\
\hline
   CI Tau & 04 33 52.014 & +22 50 30.094 & K5.5 & 4277 & $0.90\pm 0.02$ & -7.28 & -0.09 & 158.7 & (1,4,5) \\
   CY Tau & 04 17 33.728 & +28 20 46.810 & M2.3 & 3560 & $0.30\pm 0.02$ & -8.85 & -0.61 & 128.9 & (2,3,4,5)\\
   DG Tau$^*$ & 04 27 04.691 & +26 06 16.042 & K7   & 4000 & $0.70$  & -7.35 & -0.58 & 121.2 & (2,3,5,6)\\
   DL Tau & 04 33 39.077 & +25 20 38.098 & K5.5 & 4277 & $1.04\pm0.02$ & -7.62 & -0.19 & 159.3 & (1,4,5) \\
   DM Tau & 04 33 48.734 & +18 10 09.974 & M1   & 3720 & $0.55\pm 0.02$ & -7.99 & -0.82 & 145.1 & (2,3,4,6,7) \\
   DN Tau & 04 35 27.378 & +24 14 58.909 & M0.3 & 3806 & $0.87\pm0.15$ & -9.04 & -0.16 & 128.2 & (1,5,7) \\
   IQ Tau & 04 29 51.557 & +26 06 44.855 & M1.1 & 3690 & $0.74\pm 0.02$ & -8.54 & -0.67 & 131.3 & (1,4,5) \\
UZ Tau E$^*$  & 04 32 43.078 & +25 52 30.672 & M1.9 & 3574 & $1.20\pm0.03$ & -8.05 & -0.46 & 131.2 & (1,5,7) \\
\hline
\end{tabular}
\tablefoot{The $J=2000$ coordinates and distances are based on the Gaia DR2 catalog \citep{Gaia_DR2}. The DR2 distances have uncertainties of $\approx 0.7-1.2$~pc.  Spectral types, $T_{\rm eff}$ and Log$(L_*)$ for CI, DL, DN, IQ~Tau and UZ~Tau~E are adopted from \citet{Long_ea18a}. For CY, DG, and DM~Tau, the distance-corrected values are taken from \citet{Luhman_ea10} and \citet{Herczeg_Hillenbrand14}. The dynamical stellar masses for CI, CY, DL, DM and IQ~Tau were taken from \citet{Simon_ea19}. The stellar masses for DN~Tau and the UZ~Tau~E binary stars were taken from \citet{Long_ea22a}. The stellar mass of DG~Tau is taken from \citet{Gangi_ea22} (no uncertainties provided). The mass accretion rates, except for DM~Tau, are from \citet{Gangi_ea22}. The mass accretion rate for DM~Tau is taken from \citet{Ricci_ea10}.  $^*$ denotes the bynary systems. 
References are (1) \citet{Long_ea18a}, (2) \citet{Luhman_ea10}, (3) \citet{Herczeg_Hillenbrand14}, (4) \citet{Simon_ea19}, (5) \citet{Gangi_ea22}, (6) \citet{Ricci_ea10}, and (7) \citet{Long_ea22a}.
}
\end{table*}

\begin{table*}
\caption{Disk properties.\label{tab:disk_prop}}
\centering
\begin{tabular}{lllllllll}
\hline\hline
Source & V$_{\rm LSR}$  & Incl.     & P.A.  & $R_{\rm disk}$ & $M^{\rm dust}_{\rm disk} \times 100$ & $S_\nu$(1.3mm) & Age & References \\
       & (km\,s$^{-1}$) & ($\degr$) & ($\degr$)  & (au)           & ($M_\odot$)     & (mJy)   & (Myr) &      \\
\hline      
   \object{CI Tau} & 5.74 & $49.99^{+0.11}_{-0.12}$ & $281.22^{+0.13}_{-0.13}$ & $520 \pm 13$ & 0.016 & $125.3 \pm 6.2$  &  $2.5^{+2.0}_{-1.1}$ & (1,4,6) \\
   \object{CY Tau} & 7.25 & $34 \pm 3$ & $63 \pm 5$   & $295 \pm 11$ & 0.019 & $111.1 \pm 2.9$  &  $2.3^{+0.7}_{-0.6}$ & (1,3) \\
   \object{DG Tau} & 5.5 & $32 \pm 2$ & $60 \pm 4$ & $300 \pm 50$ & 0.036 & $389.9 \pm 4.6$   & $0.28^{+0.12}_{-0.08}$ & (2,3) \\
   \object{DL Tau} & 6.10 & $44.95^{+0.09}_{-0.09}$ & $322.14^{+0.15}_{-0.14}$ & $463 \pm 6$  & 0.025 & $204.4 \pm 1.9$ &  $3.5^{+2.8}_{-1.6}$ & (1,4,6)  \\
    \object{DM Tau} & 6.05 & $-34.7 \pm 0.7$ & $64.7 \pm 0.2$    & \upd{$900 \pm 10$} & 0.025 & $108.5 \pm 2.4$   & 1-5 & (1,5,7) \\
   \object{DN Tau} & 6.38 & $35.18^{+0.2}_{-0.22}$ & $349.19^{+0.36}_{-0.38}$  & $241 \pm 7$  & 0.013 & $88.61^{+0.09}_{-0.2}$ &  $0.9^{+0.6}_{-0.4}$ &  (1,4,6) \\
   \object{IQ Tau} & 5.47 & $58.2 \pm 0.2$ & $313.8 \pm 0.22$      & $220 \pm 15$ & 0.009 & $64.11^{+0.25}_{-0.34}$ &  $4.2^{+4.1}_{-2.0}$ &  (1,4,6) \\
   \object{UZ Tau E}  & 5.70 & $56.15 \pm 0.07$ & $0.39 \pm 0.08$  & $300 \pm 20$ & 0.018 & $149.9 \pm 1.4$ &  $1.3^{+1.0}_{-0.6}$ & (1,4,6) \\
\hline
\end{tabular}
\tablefoot{The systemic velocities, the outer gas radii, and the disk masses measured from the dust emission for all disks but DG~Tau are adopted from \citet{Guilloteau_ea16}, and for DG~Tau from \citet{Guedel_ea18a} and \citet{Guilloteau_ea11a}. 
The $S_\nu$(1.3mm) continuum fluxes for all disks but IQ and DN~Tau are taken from \citet{Guilloteau_ea16} and for the IQ~Tau and DN~Tau from \citet{Long_ea18a}. The inclination and positional angle (PA) for CI, DL, DN, IQ and UZ~Tau~E are taken from \citet{Long_ea18a}, for DM~Tau from \citet{Pietu_ea07}, and for CY and DG~Tau from \citet{Guilloteau_ea11a}. Position angles are those of the disk rotation axis. The disk ages  for CI, DL, DN, IQ~Tau and UZ~Tau~E are taken from \citet{Long_ea19_taurus}, for CY and DG~Tau from \citet{Guilloteau_ea11a}, and for DM~Tau from \citet{Law_ea22a}. 
References are (1) \citet{Guilloteau_ea16}, (2) \citet{Guedel_ea18a}, (3) \citet{Guilloteau_ea11a}, (4) \citet{Long_ea18a}, (5) \citet{Pietu_ea07},  (6) \citet{Long_ea19_taurus}, and (7) \citet{Law_ea22a}.}
\end{table*}

\subsection{Main science goals}
\label{sec:goals}

\upd{In this first paper, we aim to derive the disk temperature and density structures, using the optically thick and thin CO isotopologue data, and the power-law radiative transfer model DiskFit \citep[e.g.,][]{Pietu_ea07}. The best-fit radial temperature and column density profiles that are presented in this paper were also utilized as priors to reconstruct 1+1D radial and vertical disk structures using the physical, chemical and line radiative transfer model DiskCheF \citep[presented in][]{Franceschi_ea24a}.} 

\upd{In the next study, we aim to constrain} the disk ionization from a combination of the optically thick and thin HCO$^+$~(3-2) and N$_2$H$^+$~(3-2) isotopologue lines, using the best-fit disk physical structures obtained from the CO data. 
The disk turbulence (or the upper limits) will be constrained from the CS data to minimize the thermal broadening contribution to the local line profiles.

After that, using the reconstructed best-fit physical structures and a variety of detected molecules, the CHNOS chemistry, the volatile budget, and the presence of the large-scale radial chemical substructures in the disk sample will be studied. 
We will infer the gas-phase elemental ratios and depletion factors, 
and will investigate the impact of the high-energy FUV and X-ray radiation on disk chemistry from the combination of C$_2$H, CN, HCN, and HNC. 
The efficiency of the D-, $^{13}$C-, and $^{18}$O-fractionation will be inferred from the analysis of the observed ratios of DCN/HCN, DNC/HNC, N$_2$D$^+$/N$_2$H$^+$, etc., coupled with the chemical model with fractionation processes. 
The inferred chemical abundances, elemental ratios, depletion factors and isotopic ratios will be correlated with the key stellar and disk physical quantities (mass, size, gas temperature, etc.).

\section{Observations and methods}
\label{sec:obs}

We observed the GO~Tau disk in a pilot 1~mm line survey with NOEMA, using 9 antennas in C configuration ($\sim 0.9\arcsec$ beam), reaching the sensitivity of $8-12$~mJy/beam at $0.3$~km\,s$^{-1}$ resolution after $\sim 4-6$~hours of the on-source integration  (Guilloteau et al., in prep.). This combination between the resolution and integration time allowed us detecting and at least partially resolving the line emission from key CHONS species and their isotopologues, and is optimal for the line surveys in nearby protoplanetary disks. The four PolyFiX spectral settings in the 210-280~GHz range with a total 62~GHz coverage were required to target the majority of the detected or potentially detectable molecular transitions. We employed the same observing strategy for the PRODIGE project, where each disk is observed for $4-6$~hours using the same spectral setup and antenna configuration each year (four different spectral setups between 2020 and 2024). 

\subsection{PolyFiX/NOEMA observations}
\label{sec:data_red}
\upd{Observations} of the eight T~Tau disks were carried out at the NOEMA interferometer with the PolyFiX correlator in Band~3 (1.3\,mm) between 2020, January 9 and 2020, November 30 (project L19ME; PI: Thomas Henning, co-PI P. Caselli and S19AW for DN\,Tau and CI\,Tau). We used high-resolution, 62.5kHz spectral chunks to target specific emission lines over the $\sim 15.5$~GHz PolyFiX bandwidth in dual polarization (other spectral windows had a 2~MHz resolution). The first observing run covered the main CO isotopologues in order to better characterize the disk physics, which is needed for the further analysis and chemical modeling. In addition to $^{12}$CO, $^{13}$CO, C$^{18}$O and $^{13}$C$^{17}$O~(2-1), the lines of para-H$_2$CO, DCO$^+$, DCN, DNC, $^{13}$CN, cyclic C$_3$H$_2$, C$_2$D, HC$_3$N, N$_2$D$^+$, and other rare species were targeted. The spectral setup was centered at $\sim225$GHz and covered, in total, $\sim 40$ molecular lines in 27 high resolution spectral windows. The spectral setup is shown in Fig.~\ref{fig:setup1} and targeted molecular lines and their spectroscopic parameters are listed in Table~\ref{tab:line_dat_s1} in Appendix~\ref{append:sec:setup1}. In 2021-23, two other spectral setups were observed, covering CN, $^{13}$CN, cyclic C$_3$H$_2$, ortho-H$_2$CO, HCN isotopologues and HCO$^+$ isotopologues, CS, H$_2$CO, and other species, respectively (the data will be presented in a future paper). 

Observations were carried out with ten antennas in the C configuration. Each source was observed independently (no track-sharing) with one to three tracks, using the quasars J0438+300 and 0507+179 as phase calibrators. 3C84 was the bandpass calibrator in all the tracks but the last one on UZ\,Tau\,E, where 3C454.3 was used instead. To ensure good relative \upd{flux} calibration, LkHa\,101 served as a flux calibrator in all datasets but one on UZ\,Tau\,E (this last track was calibrated using the common calibrator with the first track observed on this source the day before). MWC349 and 2010+723 have been additional flux calibrators on some tracks (for more detail, see \ref{tab:summary-obs}). \upd{Absolute flux calibration uncertainties for our CO data are within 10\% or better.} Mean on-source integration time was 5.5~hours per source. The $1\sigma$ rms noise varies from source to source and is $3.7-9.2$~mJy at 0.3~km\,s$^{-1}$ resolution.

\subsection{Data reduction}
The raw data were reduced and calibrated at IRAM, using GILDAS/CLIC package (version 24sep19~02:05~cest)\footnote{\url{http://www.iram.fr/IRAMFR/GILDAS}}. For each track, we ran the CLIC reduction pipeline and, based on the astronomer on duty's notes and after data inspection, corrected for failures (via flagging) or inaccurate baseline measurements (using measurements taken after the original observation). The calibrator fluxes (and their time variations) were bootstrapped and applied to each dataset through a final round of amplitude calibration. The $uv$ tables were generated at full spectral resolution for each source and spectral window.

\subsection{Line imaging}
\label{sec:obs_line}
Self-calibration and imaging were performed with the Imager software\footnote{\url{https://imager.oasu.u-bordeaux.fr}}, \upd{using an automatic pipeline, a custom-built line catalog, and third-party Python tools. Before any processing, we accounted for the source proper motions by using the Gaia~DR2 measurements and the UV\_PROPER\_MOTION command on each UV table.} After self-calibration, the solutions were applied to the line spectra in the respective sidebands to improve their signal-to-noise ratios. To increase the S/N ratio, the line data were resampled to 0.3~km\,s$^{-1}$ spectral resolution before imaging and deconvolution. To further maximize the sensitivity, the data were cleaned using natural weighting with the Hogbom algorithm.  After that, channel and moment maps were created.
The synthesized beams are $\approx 1.1-1.5 \arcsec \times 0.5-0.7\arcsec$, for the major and minor beam axis, respectively. The corresponding positional angles are $\sim 15-20\degr$. Even in the worst case, the beam major axis is at most at 45$\degr$ from the disk minor axis, making the disks properly spatially resolved. 

To further increase the S/N of the data, we used the Kepler pixel deprojection technique presented in \citet{Teague_ea16,Yen_ea_16,Matra_ea17}.
This technique allows shifting and stacking individual spectra by taking the disk geometry, distance and Keplerian velocities into account, \upd{boosting the signal-to-noise for detected emission. It also allows recovering the emission radial profile.}
\upd{We use the implementation available through the \texttt{KEPLER} command in the Imager package. It is similar to that available through the GoFish\footnote{\url{https://github.com/richteague/gofish}} tool \citep{GoFish}. 
The Imager implementation provides the radial brightness profile distribution by slicing the data into radial bins, and for each bin, computing the average spectrum over azimuth, thereby producing a Radius-Velocity diagram. The radial profile is obtained by retaining the maximum brightness found as a function of velocity, and an error estimate is derived from the spectra. This peak brightness estimator is more reliable than the line integrated flux, as faint line wings are always difficult to properly fit to derive such flux accurately. However, in the inner part of disk, which remains spatially unresolved, the peak brightness is reduced compared to the intrinsic source brightness due to beam dilution. The method works well because geometrical parameters (including the projected Keplerian velocities) are well defined from the bright lines such as $^{12}$CO~(2-1).}
The only exception was DG~Tau due to its complex structure and the presence of the outflow and shocks, hence only the standard line imaging was done for this source.

\subsection{DiskFit model}
The radiative transfer code DiskFit was used to fit the observed $uv$-visibilities \citep[see e.g.,][]{Pietu_ea07}. 
\sgupd{Rotationally }symmetric disk structure was assumed, with power laws used to describe disk surface density, temperature, molecular column density radial profiles, etc. The disk was assumed to be vertically isothermal, with vertically uniform relative molecular abundances and Gaussian density distribution for the total gas density. The LTE excitation conditions have been assumed, and the ray tracing was used to calculate synthetic line emission. The observational data were fitted in  Fourier space by the Levenberg–Marquardt $\chi^2$ minimization, using the same $uv$ sampling in the model as in the observed data. \sgupd{The symmetric $1\sigma$ error bars that include thermal noise and phase calibration errors were obtained through the covariance matrix. This minimization method is sufficient despite the large number of model parameters (up to 17), because as discussed by \citet{Pietu_ea07}, except for the rotation velocity-inclination pair coupled through $v\sin{i}$, parameters are only weakly coupled as can be seen from the MCMC fitting done by \citet{Simon_ea17}, their Fig.B11.}

\upd{The validity of the power law assumptions in the DiskFit model relies on rapid decrease of the disk surface density with radius, the known Keplerian rotation pattern for the gas, and power-law-like radial decrease of the temperature as probed from the CO emission. As was discussed in \citet{Pietu_ea07} and \citet{Phuong_ea20}, these are reasonable assumptions for outer, $\gtrsim 50-100$~au regions in low-mass, Class~II T~Tauri disks without severe substructures, as considered in the present study. In inner, $\lesssim 30-80$~au disk regions, temperature distributions could become flatter, as was found, for example, by the modeling of the CO ALMA data from MAPS large program \citep{Law_ea22a}.} 

\upd{While these assumptions apply reasonably well for at least CO and its isotopologues in the sample of disks studied here, the assumption of a power-law radial decrease in molecular column density for other species could be less justified due to the presence of chemical and/or excitation effects, which will be investigated in subsequent papers. The azimuthally averaged CO intensity profiles presented below do not show large-scale substructures. 
We have further tested applicability of the power-law DiskFit model for recovering the underlying disk physical parameters. For that, we have generated synthetic CO isotopologue datacubes with a 2D-line radiative transfer model URANIA, assuming a power-law surface density profile, hydrostatic equilibrium in vertical direction, thermal structure computed by radiative transfer, and using a full gas-grain chemistry, and found that the DiskFit model can recover the underlying disk properties well \citep{Pavlyuchenkov_ea07}.}

To facilitate the fitting, we used the stellar masses and geometric parameters (position, inclination, orientation) from the previous studies and the CO data (see Tables \ref{tab:stellar_prop}-\ref{tab:disk_prop}). In a first iteration, a fit of the optically thick $^{12}$CO~(2-1) line was used to determine the gas temperature, assuming it follows a power law distribution ($T_k(r)= T_0\times(r/100~\mathrm{au})^{-q}$). High enough $^{12}$CO column densities were assumed to ensure optical thickness of the synthetic $^{12}$CO~(2-1) emission (e.g., $N(\textrm {CO}) > 10^{16}$~cm$^{-2}$ at 100~au). The temperature $T_k(r)$ derived from fitting the optically thick $^{12}$CO line was used to fit mostly optically thin $^{13}$CO~(2-1) data and to determine the $^{13}$CO column density distribution ($N(\mathrm{CO}; r)= N_0\times(r/100~\mathrm{au})^{-p}$). As an initial guess for the $^{13}$CO column densities, the inferred $^{12}$CO column densities scaled down by a factor of 70 were utilized. We then iterated between $^{12}$CO and $^{13}$CO data in order to optimize the fit results. Finally, to successfully fit the weakest C$^{18}$O~(2-1) emission, the stellar masses, geometric parameters, and temperature profiles had to be fixed, using the best-fit values derived from the $^{13}$CO fitting. \upd{The results of the modeling are presented below in Sec.~\ref{subsec:res_diskfit}.}

\section{Results}
\label{sec:res}

\subsection{CO line images}
\label{subsec:res:images}

\begin{figure*}
\centering
\includegraphics[width=0.72\hsize,clip]{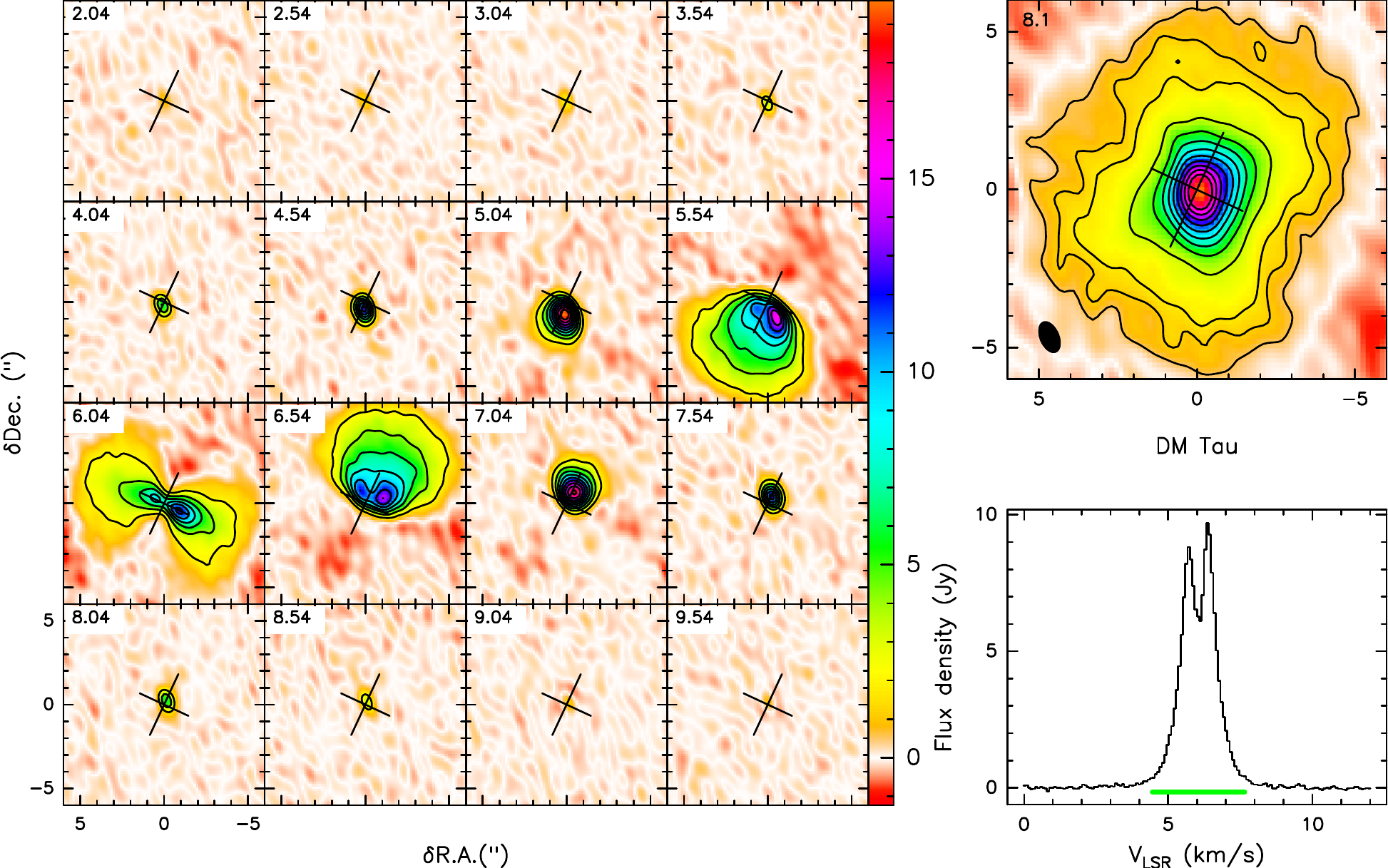}
\caption{Observations of $^{12}$CO (2-1) emission in the DM~Tau disk. Shown are the channel map of the observed line brightness distribution (left), the moment-zero map (top right), and the integrated spectrum (bottom right). The channel map shows 16 velocity channels with a step of $0.5$~km\,s$^{-1}$ in the [-4.5, +4.5]~km\,s$^{-1}$ range around the systemic velocity. The contour lines start at $2$~K, with a step of $2$~K. The color bar shows line brightness temperatures (K). The contour lines in the moment-zero plot start at 3, 6, and $9\sigma$, with a step of $6\sigma$ afterward. The $1\sigma$ rms noise (mJy\,km\,s$^{-1}$) is shown in the upper left corner of the moment-zero plot, and the synthesized beam is depicted by the dark ellipse in the left bottom corner.
}
\label{fig:DM_Tau-CO_comp}
\end{figure*}

\begin{figure*}
\sidecaption  \includegraphics[width=12cm,clip]{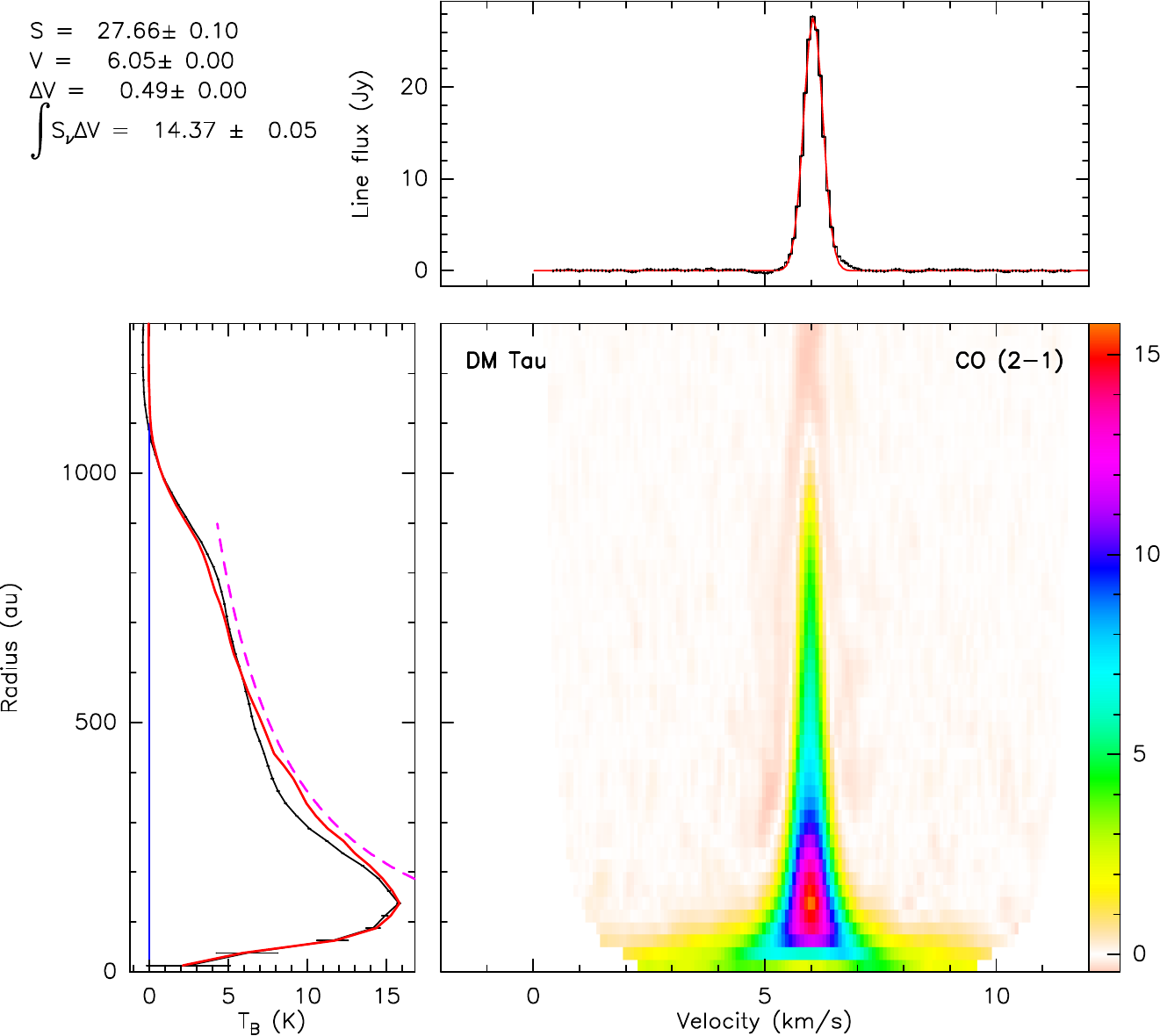}
\caption{Observations of $^{12}$CO (2-1) emission in the DM~Tau disk. Shown is a pixel-deprojected Keplerian plot consisting of the three panels: (left) the radial profile of the line brightness temperature (K) \upd{with observations and errorbars in black, and profile derived from the best-fit disk model in red. The magenta dashed curve indicates the expected brightness temperature of an optically thick line derived from temperature power law}. (Top) the integrated spectrum (black line) overlaid with the best-fit Gaussian profile (red line). (Bottom right) aligned and stacked line intensity (K) as a function of disk radius (Y-axis; au) and velocity (X-axis; km\,s$^{-1}$). The color bar units are in Kelvin.}
\label{fig:DM_Tau-CO_kepl}
\end{figure*}

We present the $^{12}$CO~(2-1) moment map, channel maps, integrated spectrum, and Keplerian plot for the bright and large DM~Tau disk in Figs.~\ref{fig:DM_Tau-CO_comp}-\ref{fig:DM_Tau-CO_kepl}. These plots look overall similar for the other CO isotopologues and for the other disks (\upd{although, usually, with more noise and, in some cases, contamination by molecular cloud}, see Appendix~\ref{append:sec:obs_plots}). The $^{12}$CO channel map in Fig.~\ref{fig:DM_Tau-CO_comp} shows a ``butterfly'' pattern characteristic of rotating Keplerian disks. The DM~Tau $^{12}$CO observations are not affected by the foreground cloud absorption, which is also true for DN~Tau and IQ~Tau (see Appendix~\ref{append:sec:obs_plots}). In contrast, the optically thick emission lines in CI, CY, DG, DL~Tau, and UZ~Tau E are partially obscured at $\sim 4-6$~km\,s$^{-1}$. This absorption leads to asymmetric emission distributions in the integrated intensity maps and skewed double-peaked integrated spectra. For example, the NW half of the CI~Tau disk appears much less pronounced than its SE half on the CI~Tau integrated intensity map (see Fig.~\ref{fig:CI_Tau-CO_comp}). 

The Keplerian deprojection transforms the double-peaked $^{12}$CO~(2-1) integrated spectrum of the DM~Tau disk with a $\sim 2$~km\,s$^{-1}$ FWHM width and a peak value of $\sim 9.5$~Jy into a single peaked spectrum with a FWHM width of $\sim 0.5$~km\,s$^{-1}$ and a peak value of $\sim 27.7$~Jy (see  Fig.~\ref{fig:DM_Tau-CO_kepl}, top panel). This single-peaked integrated spectrum is well-fitted by a single Gaussian. The ``teardrop'' pattern visible in the bottom right panel of Fig.~\ref{fig:DM_Tau-CO_kepl} is a characteristic signature of molecular emission from a Keplerian disk \citep[e.g.,][]{Teague_ea16}. It is fully symmetric with respect to the systemic velocity of $6.05$~km\,s$^{-1}$. The high-velocity ``halo'' visible at the inner $\lesssim 100$~au radii in the teardrop plot represents the CO emission from the innermost disk region with the strongest gradients of physical conditions. It falls within a single beam and hence suffers from beam smearing and dilution.

\upd{The radial profile of the $^{12}$CO~(2-1) line brightness temperature depicted in red in left panel in Fig.~\ref{fig:DM_Tau-CO_kepl} is derived from the best-fit disk model, while the magenta dashed line shows the expected brightness of a completely optically thick line. The overall agreement between these curves and the observed profile indicates that CO is mostly optically thick and that the power-law approximation for the temperature is appropriate at this scale. The line brightness temperature distribution has a skewed Gaussian-like profile, with an inner ``hole’’ inside $r \sim 150$~au, followed by an extended bell-like peak with $T_\mathrm{B} \sim 16$~K, and a steady decline until $r \sim 1,000$~au. The apparent drop within the inner $r \lesssim 150$~au is a result of beam dilution.} 
The radial extent of the $^{12}$CO~(2-1) line brightness temperature of $\sim 1,000$~au is larger than the CO gas radius estimate of $\sim 900 \pm 10$~au (Table~\ref{tab:disk_prop}) \upd{due to our limited angular resolution.} 

The line brightness radial profiles look similar in the other disks and for the other CO isotopologue lines, and do not show any significant substructures. The targeted disks do not have prominent inner holes with sizes of $\sim 50-100$~au, neither in the dust nor gas distributions, and the $\sim 100$~au spatial resolution of our data does not permit us to infer the presence of smaller substructures in the disks. The typical peak $T_\mathrm{B}$ values are $\sim 6.5-16$~K for $^{12}$CO~(2-1), $1.2-7.5$~K for $^{13}$CO~(2-1), and $0.2-1.8$~K for C$^{18}$O~(2-1), respectively. The $\approx 0.8-0.9\arcsec$ beam ($\sim 100-150$~au spatial scale) leads to the radially smooth $T_\mathrm{B}(r)$ profiles, without any significant substructures visible in the plots. The emission radii vary from disk to disk and differ for each CO isotopologue, with the most extended emission in the $^{12}$CO~(2-1) lines. \upd{As noted above, the inner ``holes'' with radii $\lesssim 100-150$~au in the brightness temperature profiles are due to the beam dilution}.  

\upd{While the analysis of the large disk around DM~Tau shows that the DiskFit modeling in terms of power laws is reasonably appropriate, the parameters derived from such models do start suffering from unavoidable degeneracy when the disk radius is small due to the limited angular resolution. For the smaller disks (DN\,Tau, IQ\,Tau and UZ\,Tau~E), deriving both the column density and the temperature law becomes impossible as the optically thick core cannot be well separated from the optically thin outer part. The geometric parameters are, however, unaffected. Temperature can be derived from $^{12}$CO, while $^{13}$CO and C$^{18}$O column densities can be derived assuming the same temperature as given from $^{12}$CO. Furthermore, in the relevant temperature range (20-30 K), the derived column densities are rather insensitive  to the temperature. We provide below more} specific details for each source.

\paragraph{CI~Tau:}
Both the moment-zero map and the integrated spectrum of the $^{12}$CO~(2-1) emission are affected by the foreground absorption at \upd{4.2-6.5}\,km\,s$^{-1}$ (\ref{fig:CI_Tau-CO_comp}). It also causes the asymmetry of the teardrop plot and the slightly skewed deprojected integrated spectrum that cannot be fitted with a single Gaussian (Fig.~\ref{fig:CI_Tau-CO_kepl}). The radial profile of the line brightness temperature is \upd{fainter than the best fit profile because of the absorption, and extends only up to $\sim 600-650$~au}. In contrast, the $^{13}$CO and C$^{18}$O~(2-1) data unaffected by the cloud show smooth profiles and a broader extent of the line brightness temperature up to radius of $\sim 900$~au (Figs.~\ref{fig:CI_Tau-13CO_comp}-\ref{fig:CI_Tau-C18O_kepl}).

\paragraph{CY~Tau:}
The $^{12}$CO (2-1) emission is also affected by the foreground absorption that is visible as stripes in the channel map at $V_\mathrm{LSR}\sim 4.5$ and $5$~km\,s$^{-1}$ (\ref{fig:CY_Tau-CO_comp}). This is far from the systemic velocity of $7.3$~km\,s$^{-1}$, and thus both the moment-zero map and integrated spectrum are not strongly affected by the absorption. The asymmetry of the teardrop plot toward the redshifted wing at radii $\lesssim 100$~au is \upd{due to the cloud contamination} (Fig.~\ref{fig:CY_Tau-CO_kepl}). The CO emission for all isotopologues is compact, $R_\mathrm{CO} \sim 300$~au. \upd{Deriving a temperature profile from the CO data is ambiguous because of the combination of cloud contamination and relatively compact ($R_\mathrm{CO} \sim 300$~au) disk.}

\paragraph{DG~Tau:}
The $^{12}$CO~(2-1) emission in the channel map reveals a complex structure, which consists of a large-scale envelope with filamentary structure, a bow shock at velocities of $\sim 2.5-4$~km\,s$^{-1}$, and a Keplerian disk (Fig.~\ref{fig:DG_Tau-CO_comp}). The large-scale outflow emission is not directly visible on the $^{12}$CO~(2-1) map, partly due to the lack of short spacing data. Strong foreground cloud and envelope absorption lead to the multiple absorption dips in the integrated spectrum. The same complex structure is present in more optically thin $^{13}$CO and C$^{18}$O~(2-1) emission maps (Figs.~\ref{fig:DG_Tau-13CO_comp}-\ref{fig:DG_Tau-C18O_comp}). \upd{The C$^{18}$O emission indicates a possible small ($\lesssim 100$\,au radius) disk, but with no obvious signature for Keplerian rotation, given the weak S/N.}

\paragraph{DL~Tau:}
The $^{12}$CO (2-1) emission shows strong foreground absorption at $V_\mathrm{LSR} \sim 5-6.5$~km\,s$^{-1}$, which is visible in the moment-0th map and integrated spectrum (Fig.~\ref{fig:DL_Tau-CO_comp}). Even more optically thin $^{13}$CO~(2-1) emission is affected, and shows an arc-like integrated intensity distribution that is located off the center (Fig.~\ref{fig:DL_Tau-13CO_comp}, top right). The C$^{18}$O~(2-1) emission is weak and barely visible in the Keplerian deprojected plot (Fig.~\ref{fig:DL_Tau-C18O_kepl}). The CO line brightness temperatures are low in this disk, in particular for optically thin $^{13}$CO and C$^{18}$O~(2-1) lines. The peak $T_\mathrm{B}$ values are $\sim 7.2$~K for $^{12}$CO~(2-1), $\sim 1.3$~K for $^{13}$CO~(2-1), and only $\sim 0.2$~K for C$^{18}$O~(2-1), respectively. \upd{The observed $^{12}$CO~(2-1) radial brightness profile differs from the fitted one because of cloud contamination around the systemic velocity.} \upd{The $^{12}$CO~(2-1) emission is clearly optically thick, and declines more steeply at the disk edge than for the minor CO isotopologues.}

\paragraph{DM~Tau:}
The $^{12}$CO (2-1) emission is not affected by the foreground absorption. The corresponding line brightness distribution $T_\mathrm{B}(r)$ has an inner hole inside $r \sim 130$~au, followed by a peak brightness, and a 3-step decline until the radius of $ \sim 1\,050$~au, with the two turn-over points at $r ~\sim 340$~au and $\sim 700$~au, respectively (Fig.~\ref{fig:DM_Tau-CO_kepl}). The steeper profile between $r \sim 800$~au and  $\sim 1\,000$~au could be caused by the tapered surface density distribution at the disk edge. The C$^{18}$O~(2-1) emission looks noisy because the DM~Tau observations were done under suboptimal conditions during late April, leading to higher noise compared to the other disks in the sample (Fig.~\ref{fig:DM_Tau-C18O_comp}).

\paragraph{DN~Tau and IQ~Tau:} 
The CO isotopologue emission is compact, $R_\mathrm{CO} \sim 200-300$~au (Figs.~\ref{fig:DN_Tau-CO_comp}-\ref{fig:IQ_Tau-C18O_kepl}). The C$^{18}$O~(2-1) emission is weak. The IQ~Tau disk shows low line brightness temperatures that are similar to those in the DL~Tau disk. The peak $T_\mathrm{B}$ temperatures are $\sim 9.5$~K for $^{12}$CO~(2-1), $\sim 1.2$~K for $^{13}$CO~(2-1), and only $\sim 0.2$~K for C$^{18}$O~(2-1), respectively. The FWHM widths of the CO deprojected spectra in the IQ~Tau disk are $\sim 1.5$~km\,s$^{-1}$.
For DN~Tau, the DiskFit analysis requires a central hole of 15-20~au (or substantially fainter emission within this radius) in all CO isotopologues to best represent the emission. An alternate solution is a rather flat temperature distribution (exponent $q\approx 0$), which is less physically realistic. Both solutions result in a relative lack of emission at high velocities and can explain the faintness of the line wings.

\paragraph{UZ~Tau~E:}
The CO isotopologue emission maps show the presence of the prominent disk around the spectroscopic binary in the center, accompanied by the compact binary system UZ~Tau~W (Figs.~\ref{fig:UZ_Tau-CO_comp}-\ref{fig:UZ_Tau-C18O_kepl}). \upd{Emission is also visible toward UZ~Tau~W in all CO isotopologues. The disk around UZ~Tau~E is clearly asymmetric in $^{12}$CO~(2-1), with a tail of emission opposite to the position of the UZ~Tau~W binary. It extends to a similar distance as the apparent E-W star separation, clearly suggesting that all these stars were born in a common disk. The best-fit DiskFit analysis results in a compact CO disk with an outer radius of 200~au, which does not represent the extended emission at all. Moreover, there is a strong degeneracy between the assumed CO column density profile and the derived kinetic temperature profile, which is further influenced by the non-represented outer disk emission. High-temperature solutions (37~K at 100~au, with a steep exponent of $-0.85$) are possible for $^{12}$CO, but require lower column densities ($<10^{17}$~cm$^{-2}$) and relatively flat distribution with an exponent of 0.8). The emission from the minor CO isotopologues is more compact, with an outer radius around 140~au.}

\subsection{Fluxes}
\label{subsec:res:fluxes}

\begin{table*}
\caption{Parameters of the observed CO isotopologue lines.}
\label{tab: line_fluxes_sp1}
\centering
\begin{tabular}{lllllll}
\hline\hline
Molecule & Transition & Source & $V_{\rm LSR}$ & $\Delta V_{\rm FWHM}$ & $S_{\rm peak}$ & $\int S_{\nu} \Delta V$ \\
\hline
          &           &        & (km/s)        & (km/s)                & (Jy)           &  (Jy~km~s$^{-1}$)  \\
\hline
CO & (2-1) & CI~Tau &        $5.88\pm0.004$  & $0.97\pm0.01$   & $4.35\pm0.04$  & $4.48\pm0.04$ \\
& & CY~Tau &                 $7.26\pm0.01$ & $0.70\pm0.01$   & $2.94\pm0.04$  & $2.18\pm0.03$ \\
& & DL~Tau &                 $6.10\pm0.01$ & $0.70\pm0.01$   & $4.43\pm0.06$  & $3.30\pm0.05$ \\
& & DM~Tau &                 $6.05\pm0.004$ & $0.49\pm0.004$ & $27.66\pm0.10$ & $14.37\pm0.05$ \\
& & DN~Tau &                 $6.41\pm0.01$ & $0.93\pm0.02$   & $3.18\pm0.05$  & $3.16\pm0.05$ \\
& & IQ~Tau &                 $5.53\pm0.01$ & $1.47\pm0.02$   & $2.26\pm0.03$  & $3.53\pm0.04$ \\
& & UZ~Tau~E &               $5.85\pm0.01$  & $1.57\pm0.01$   & $4.97\pm0.04$  & $8.31\pm0.07$ \\
\hline 
$^{13}$CO & (2-1) & CI~Tau & $5.76\pm0.004$ & $0.66\pm0.01$ & $5.32\pm0.06$  & $3.75\pm0.04$ \\
&  & CY~Tau &                $7.26\pm0.01$ & $0.65\pm0.02$   & $1.20\pm0.03$  & $0.83\pm0.02$ \\
&  & DL~Tau &                $6.09\pm0.01$   & $0.53\pm0.03$   & $1.01\pm0.04$  & $0.56\pm0.02$ \\
&  & DM~Tau &                $6.04\pm0.004$  & $0.44\pm0.01$ & $9.63\pm0.13$  & $4.53\pm0.06$ \\
&  & DN~Tau &                $6.41\pm0.01$ & $0.93\pm0.02$   & $0.62\pm0.01$  & $0.61 \pm 0.01$ \\
&  & IQ~Tau &                $5.55\pm0.02$   & $1.31\pm0.05$   & $0.21\pm0.01$& $0.29\pm0.01$ \\
&  & UZ~Tau~E              & $5.72\pm0.01$   & $2.23\pm0.01$   & $0.50\pm0.004$& $1.18\pm0.01$ \\
\hline 
C$^{18}$O & (2-1) & CI~Tau & $5.74\pm0.01$   & $0.83\pm0.03$   & $0.95\pm0.03$  & $0.84\pm0.03$ \\
 &  & CY~Tau &               $7.27\pm0.02$   & $0.70\pm0.05$   & $0.30\pm0.02$  & $0.22\pm0.01$ \\
 &  & DL~Tau &               $6.23\pm0.03$   & $0.63\pm0.07$   & $0.13\pm0.01$  & $0.08\pm0.01$ \\
 &  & DM~Tau &               $6.05\pm0.01$   & $0.40\pm0.02$   & $1.77\pm0.06$  & $0.76\pm0.03$ \\
 &  & DN~Tau &               $6.49\pm0.08$   & $1.48\pm0.19$   & $0.06\pm0.01$ & $0.09\pm0.01$ \\
 &  & IQ~Tau &               $7.40\pm0.10$   & $1.56\pm0.24$   & $0.040\pm0.01$& $0.07\pm0.01$ \\
 &  & UZ~Tau~E &             $5.61\pm0.02$   & $2.26\pm0.04$   & $0.11\pm0.004$& $0.27\pm0.01$ \\
\hline
\end{tabular}
\tablefoot{The observed values along with the $1 \sigma$ uncertainties obtained from the Gaussian fitting of the Kepler-deprojected integrated spectrum.
Column 4: LSR velocity. Column 5: Full width half maximum line width. Column 6: Peak flux density. Column 7: Integrated flux (line area).}
\end{table*}

The observed peak and integrated CO fluxes, along with the systemic velocities and the line widths, are presented in Table~\ref{tab: line_fluxes_sp1}. \upd{The $1\sigma$ flux uncertainties were estimated from the rms noise in the emission-free channels.} 
The system velocities have been derived with a precision of $\approx 4 - 100$~m\,s$^{-1}$ (with the uncertainties stemming from the Gaussian fit of the spectra). They vary between about 5.5 and 7.4\,km~s$^{-1}$, depending on the source, which are typical values for the Taurus disks. The line widths corrected for the disk orientation, but not for Keplerian rotation for $^{12}$CO~(2-1) are $0.49-1.58$~km~s$^{-1}$, for $^{13}$CO~(2-1) are $0.45 - 2.24$~km~s$^{-1}$, and for C$^{18}$O~(2-1) are $0.40-2.31$~km~s$^{-1}$. The lines are the widest for the disks around IQ~Tau and UZ~Tau~E, \upd{which are the smallest disks and} have the highest inclination angles above $55\degr$, plus their C$^{18}$O data have low S/N and noisy.

\begin{figure}
\centering
\includegraphics[width=0.96\hsize,clip]{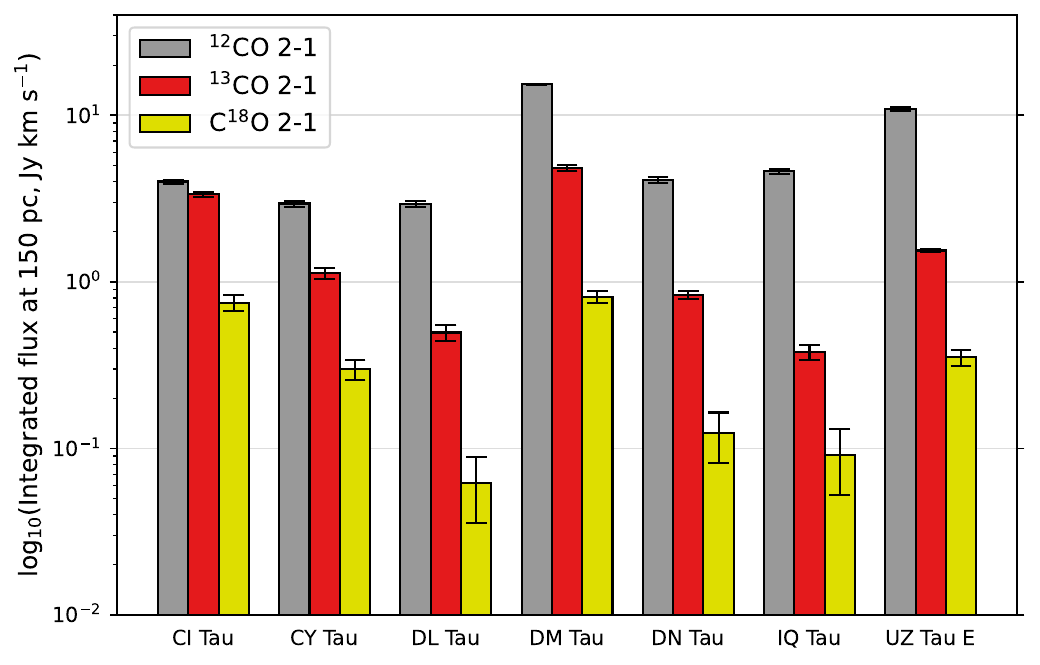}
\caption{Observed CO-integrated fluxes rescaled to the same distance of 150~pc. The $\pm 3 \sigma$ rms noise is depicted by the vertical lines with caps. The Y-axis has a $\log$ scale.}
\label{fig:int_fluxes}
\end{figure}

\begin{figure}
\centering
\includegraphics[width=0.96\hsize,clip]{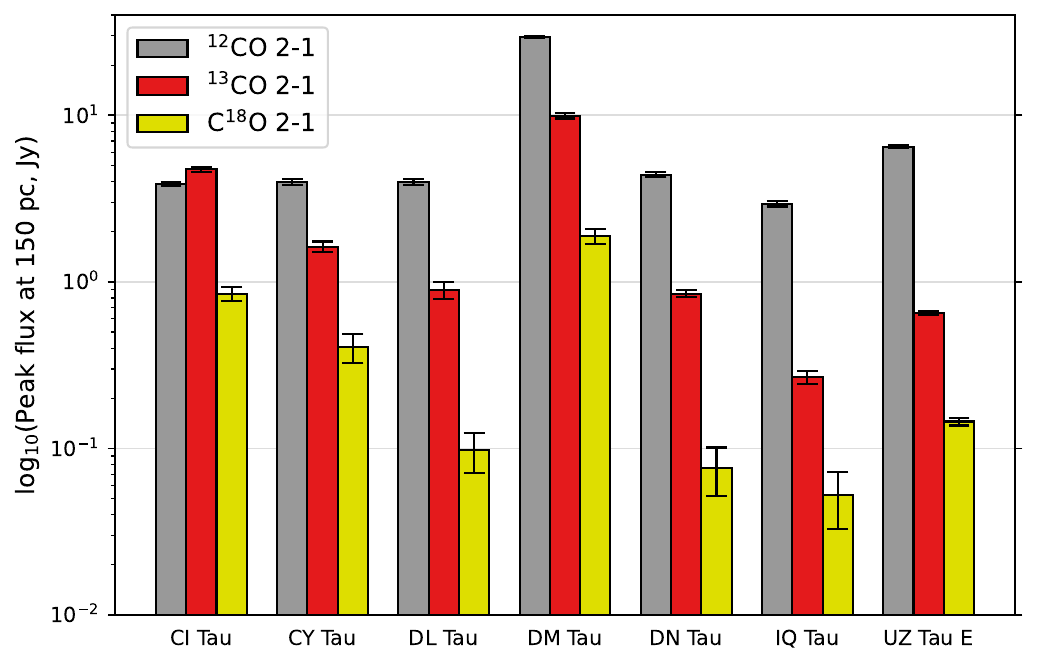}
\caption{Same as Fig.~\ref{fig:int_fluxes} but for the peak fluxes.}
\label{fig:peak_fluxes}
\end{figure}

The integrated fluxes vary between 2.94 and 15.35\,Jy~km~s$^{-1}$ for $^{12}$CO~(2-1), 4.82 and 0.38\,Jy~km~s$^{-1}$ for $^{13}$CO~(2-1), and 0.81 and 0.06\,Jy~km~s$^{-1}$ for C$^{18}$O~(2-1), respectively. Similarly, the peak fluxes (flux densities) vary between 2.95 and 29.54\,Jy for $^{12}$CO~(2-1), 9.95 and 0.27\,Jy for $^{13}$CO~(2-1), and 1.89 and 0.05\,Jy for C$^{18}$O~(2-1), respectively. The observed peak and integrated fluxes rescaled to the same 150~pc distance, \upd{as well as their uncertainties estimated from the rms noise in the emission-free channels} are shown in Figs.~\ref{fig:int_fluxes}-\ref{fig:peak_fluxes}.

The large ($\sim 800$~au) and massive ($\sim 0.01-0.045 M_{\odot}$) disk around DM~Tau is the brightest disk in our sample, with the $^{13}$CO and C$^{18}$O~(2-1) fluxes that are comparable with the $^{12}$CO and $^{13}$CO~(2-1) fluxes from the other disks. The next brightest disk is CI~Tau (with partly blocked $^{12}$CO~(2-1) emission), followed by the CY~Tau and UZ~Tau~E disks, and then by the rest of the sample. The weakest optically thin $^{13}$CO and C$^{18}$O~(2-1) emission comes from DL~Tau, DN~Tau and IQ~Tau. The $\sim 500$~au DL~Tau disk is twice as large as  the compact, $\sim 200-250$~au DN~Tau and IQ~Tau disks (Table~\ref{tab:disk_prop}). Moreover, the DL~Tau disk is about $\sim 2-3$ times more massive ($\sim 0.025 M_\odot$) than the DN~Tau ($\sim 0.013 M_\odot$) and IQ~Tau ($\sim 0.009 M_\odot$) disks, which have the lowest masses in the sample. Thus, the low $^{13}$CO and C$^{18}$O~(2-1) fluxes and the low line brightness temperatures in the DL~Tau disk indicate that either the DL~Tau disk mass has been previously overestimated or that CO is severely depleted in this system (see below). The low $^{13}$CO and C$^{18}$O~(2-1) fluxes as well as line brightness temperatures in DN~Tau and IQ~Tau disks are likely due to their low disk masses.

\upd{The integrated fluxes obtained with NOEMA are equal to within the absolute flux calibration accuracy} with the previous observations. For example, \upd{\citet{Rota_ea22} have measured the integrated $^{12}$CO and $^{13}$CO~(2-1) fluxes with ALMA in UZ~Tau~E, and found values of $7.58 \pm0.1$ and $1.33 \pm 0.04$~Jy\,km/s, respectively. 
These values agree well with the integrated $^{12}$CO and $^{13}$CO~(2-1) fluxes of $8.31 \pm 0.07$ and $1.18 \pm 0.01$~Jy\,km/s measured in this study.} \citet{Long_ea22a} have used ALMA and SMA and derived the integrated $^{12}$CO~(2-1) fluxes for DL~Tau, DM~Tau and UZ~Tau~E to be 7.05, 15.21, and 7.32\,Jy~km~s$^{-1}$, respectively. Our values are 3.32, 14.37, and 8.32\,Jy~km~s$^{-1}$, respectively. The discrepancy between the NOEMA and ALMA/SMA data is $\sim 10-15 \%$ for DM~Tau and UZ~Tau~E, and a factor of 2 for DL~Tau. The reason for such a \upd{ discrepancy for the heavily cloud-contaminated DL~Tau system is most likely due to a different degree of filtering of the extended emission. The} integrated $^{13}$CO~(2-1) fluxes in DL~Tau, DM~Tau, DN~Tau, IQ~Tau and UZ~Tau~E measured by \citet{Guilloteau_ea13} with the IRAM 30-m antenna agree with our $^{13}$CO~(2-1) fluxes within a range of $\sim 10\%$ - a factor of 2. Finally, our integrated $^{12}$CO, $^{13}$CO, and C$^{18}$O~(2-1) fluxes from CI~Tau, CY~Tau, DL~Tau, and IQ~Tau agree within \upd{$\sim 10-80\%$} with the SMA observations of \citet{Williams_Best14}.

\section{Discussion}
\label{sec:diss}

\subsection{Best-fit disk parameters derived with DiskFit}
\label{subsec:res_diskfit}

\subsubsection{Stellar properties and disk geometries}
\label{subsubsec:diskfit_geom}

\begin{table*}
\caption{
\label{tab:diskfit_geom}
Best-fit parameters of the stellar properties and disk geometry.}
\centering
\begin{tabular}{llllllllllllll}
\hline\hline
Source & Line &  $V_\mathrm{LSR}$ & $M_{*}$ & PA & Inclination & Gas radius \\
       &      &  (km/s)            & (M$_\sun$)   & (deg) & (deg) & (au) \\
\hline
CI~Tau & $^{12}$CO (2-1) &  $  5.78\pm   0.01$ & $ 1.035$ & $ 281.7\pm  0.1$ & $  48.4\pm  0.2$ & $   518\pm 1$  \\ 
 & $^{13}$CO (2-1) &  $  5.74\pm   0.01$ & $ 0.973$ & $ 281.7\pm  0.1$ & $  46.6\pm  0.9$ & $   517 \pm 14$  \\ 
 & C$^{18}$O (2-1) &  [$5.73$] & [$0.965$] & [$282.0$] & [$  46.9$] & [$   511$]  \\ 
\hline
CY~Tau & $^{12}$CO (2-1) &  $  7.31\pm   0.01$ & $ 0.543$ & $  64.7\pm  0.2$ & $  27.8\pm  3.9$ & $   251\pm 1$  \\ 
 & $^{13}$CO (2-1) &  $  7.31\pm   0.01$ & $ 0.492$ & $  63.9\pm  0.2$ & $  25.5\pm  0.2$ & $   220\pm 5$  \\ 
 & C$^{18}$O (2-1) &  [$  7.27$] & [$ 0.522$] & [$  65.0$] & [$  28.1$] & [$   214$]  \\ 

\hline
DL~Tau & $^{12}$CO (2-1) &  $  6.05$ & $ 1.283$ & $ 320.4\pm  0.2$ & $  36.7\pm 18.1$ & $   621\pm 17$  \\ 
 & $^{13}$CO (2-1) &  $  6.04\pm   0.01$ & $ 1.047$ & $ 320.2\pm  0.7$ & $  44.7\pm  4.8$ & $   499\pm 16$  \\ 
 & C$^{18}$O (2-1) &  [$  6.04$] & [$ 1.047$] & [$ 320.2$] & [$  44.7$] & [$   499$]  \\ 

\hline
DM~Tau & $^{12}$CO (2-1) &  $  6.05\pm   0.01$ & $ 0.496$ & $  65.5\pm  0.4$ & $ -33.7\pm  0.2$ & \textsg{$900 \pm 10$}  \\ 
 & $^{13}$CO (2-1) &   $  6.02\pm   0.01$ & $ 0.515$ & $  65.9\pm  0.2$ & $ -34.8\pm  7.5$ & $   688\pm 9$  \\ 
 & C$^{18}$O (2-1) &   $  6.04\pm   0.01$ & $ 0.501$ & $  66.3\pm  1.4$ &[$ -36.0$] & $   530\pm 10$  \\ 

\hline
DN~Tau & $^{12}$CO (2-1)&  $  6.40\pm   0.00$ & $ 0.756$ & $ 172.0\pm  0.2$ & $  32.7\pm  0.7$ & $   287\pm 2$  \\ 
 & $^{13}$CO (2-1)&  $  6.37\pm   0.03$ & $ 0.657$ & $ 170.9\pm  0.8$ & $  36.3\pm  1.6$ & $   168\pm 2$  \\ 
 & C$^{18}$O (2-1)&  [$  6.37$] & [$ 0.657$] & [$ 170.9$] & [$  36.3$] & $   168\pm 2$  \\ 

\hline
IQ~Tau & $^{12}$CO (2-1) &  $  5.51\pm   0.02$ & $ 0.639$ & $ 311.2\pm  0.2$ & $  62.2\pm  0.6$ & $   203\pm 4$  \\ 
 & $^{13}$CO (2-1) &  $  5.54\pm   0.11$ & $ 0.658$ & $ 311.8\pm  2.2$ & $  60.0\pm  3.7$ & $   207\pm 11$  \\ 
 & C$^{18}$O (2-1) &  [$  5.54$] & [$ 0.658$] & [$ 311.8$] & [$  60.0$] & $   207\pm 11$  \\ 

\hline
UZ~Tau~E & $^{12}$CO (2-1) &  $  5.83\pm   0.01$ & $ 1.298$ & $  -2.9\pm  0.1$ & $  56.8\pm  1.3$ & $   212\pm 1$  \\ 
 & $^{13}$CO (2-1) &  $  5.77\pm   0.04$ & $ 1.294$ & $  -6.6\pm  0.2$ & $  60.5\pm  4.1$ & $   190\pm 1$  \\ 
 & C$^{18}$O (2-1) &  [$5.77$] & [$ 1.294$] & [$  -6.6$] & [$  60.5$] & [$   190$]  \\ 

\hline
\end{tabular}
\tablefoot{The best-fit values along with the $1\sigma$ uncertainties obtained by the DiskFit modeling. The values in square brackets have been kept fixed during fitting. The distances from Table~\ref{tab:stellar_prop} have been used. Column 3: Systemic velocity. Column 4: stellar mass. Column 6: Positional angle (PA). }
\end{table*}

We used the DiskFit model and obtained the best-fit parameters for the seven disks (excluding the complex DG~Tau system). The best-fit geometric parameters are presented in Table~\ref{tab:diskfit_geom}. To fit weak C$^{18}$O~(2-1) emission, \upd{the stellar masses, geometric parameters, and temperature profiles were fixed to the best-fit $^{13}$CO values (they are indicated by the square brackets in Table~\ref{tab:diskfit_geom})}. As can be clearly seen, for each disk the systemic velocities derived from the CO isotopologues agree with each other, as well as with the literature values. \upd{The accuracy of the systemic velocities derived from the $^{12}$CO and $^{13}$CO lines varies between $4$ and $20$~m\,s$^{-1}$} (Table~\ref{tab:disk_prop}). The agreement with the best-fit values derived by the Gaussian fitting of the \upd{Kepler-integrated spectra is in general within $\lesssim 20$~m\,s$^{-1}$ for all cases, except for the disks affected by the strong cloud confusion (where Kepler-deprojected spectrum is biased), and the very weak C$^{18}$O~(2-1) emission of IQ~Tau}. 

The best-fit inclination and positional angles agree with the literature values, and are within  $\lesssim 1-4\degr$ for the $^{12}$CO and $^{13}$CO data for all the sources but CY~Tau (inclination) and DN~Tau (PA), see Table~\ref{tab:disk_prop}. The PA and inclination derived from the $^{12}$CO data in DN~Tau are larger than those obtained from the $^{13}$CO data by $\approx 2$ and $3.6\degr$, respectively. For DL~Tau, the  $^{12}$CO~(2-1)-based inclination angle is smaller than the $^{13}$CO~(2-1)-based value by $\approx 8\degr$, but it has a large uncertainty of $\sim 18\degr$ due to the asymmetry of emission caused by the foreground absorption. 

The discrepancy in the best-fit orientation and the inclination of the CO emitting layers affects the best-fit dynamical stellar masses. This leads to the difference of $\lesssim 5-10\%$ in the inferred masses of CI~Tau, CY~Tau, DM~Tau, IQ~Tau and UZ~Tau~E. The best-fit stellar masses of DL~Tau and DN~Tau derived from $^{12}$CO~(2-1) are larger by $\sim 15-20\%$ than those obtained from $^{13}$CO~(2-1), see Table~\ref{tab:diskfit_geom}. 

Furthermore, the best-fit stellar masses agree within $\sim 10\%$ with the literature values (Table~\ref{tab:stellar_prop}) for CI~Tau, DL~Tau (disregarding the $^{12}$CO estimate), DM~Tau, IQ~Tau and UZ~Tau~E. The best-fit dynamical mass of DN~Tau, $\approx 0.66-0.76 M_{\odot}$ agrees with the literature value of $0.87\pm 0.15 M_{\odot}$ within the uncertainties. The strongest disagreement between the best-fit mass derived with DiskFit ($0.49-0.54 M_{\odot}$) and the dynamical value from the literature \citep[$0.30 \pm 0.02 M_{\odot}$][]{Simon_ea17} is found for the CY~Tau disk. In contrast to our $\approx 0.9\arcsec$ observations with NOEMA, the CY~Tau mass in \citet{Simon_ea17} has been derived by the DiskFit modeling of the ALMA $^{12}$CO~(2-1) data with a higher $\approx 0.23\arcsec$ resolution. Given a low inclination of $\lesssim 30\degr$ and a small disk size of $\sim 200-250$~au, it becomes difficult to precisely constrain the inclination from the CO data when the angular resolution is only twice better than the disk radius. In this case, the DiskFit fitting of the NOEMA data resulted in $i \sim 24-28\degr$, while for the \citet{Simon_ea17} ALMA data the DiskFit converged to $i \sim 30 \pm 2\degr$. The strong absorption toward CY~Tau further complicates the fitting of the NOEMA $^{12}$CO data.

Finally, the best-fit CO gas radii and the literature values shown in Table~\ref{tab:disk_prop} agree within $\sim 10-20\%$ for the entire sample. The smallest CO emission sizes are found for the DN~Tau ($\sim 170-290$~au), UZ~Tau~E ($\sim 190-210$~au), and IQ~Tau ($\sim 200-210$~au) systems. The DL~Tau and DM~Tau disks are the largest, with $R_\mathrm{CO} \sim 500-800$~au, which could be explained by their high initial sizes followed by efficient viscous spreading during their evolution \citep[e.g.,][]{1973A&A....24..337S,Lynden-BellPringle74,Manara_ea_PPVII_2022}. For CI~Tau, CY~Tau, IQ~Tau and UZ~Tau~E, the difference between various best-fit $R_\mathrm{CO}$ is $\lesssim 40$~au. In contrast, for DL~Tau, DM~Tau, and DN~Tau, the disk sizes derived from $^{12}$CO~(2-1) differ significantly from the sizes inferred from the less bright CO isotopologue lines (by $\gtrsim 100$~au). Besides the cloud absorption affecting the observed $^{12}$CO emission and the noise affecting minor CO isotopologue data, another reason could be the structure of the disk, namely, the steepness of the surface density profile and the degree of tapering toward the outer edge. For strongly tapered disks, the $R_\mathrm{CO}$ sizes should be less affected by the S/N of the respective CO data, as well as the photodissociation of CO and opacity effects. Interestingly, in the study of \citet{Guilloteau_ea11a}, it has been found from the continuum data modeling that the disks around DM~Tau and DL~Tau should be less tapered than the disks around CI~Tau, CY~Tau, and UZ~Tau~E, supporting our NOEMA observations.

\subsubsection{CO temperatures, column densities, isotopic ratios, and disk gas masses}
\label{subsubsec:diskfit_phys}

\begin{figure}
\centering
\includegraphics[width=0.96\hsize,clip]{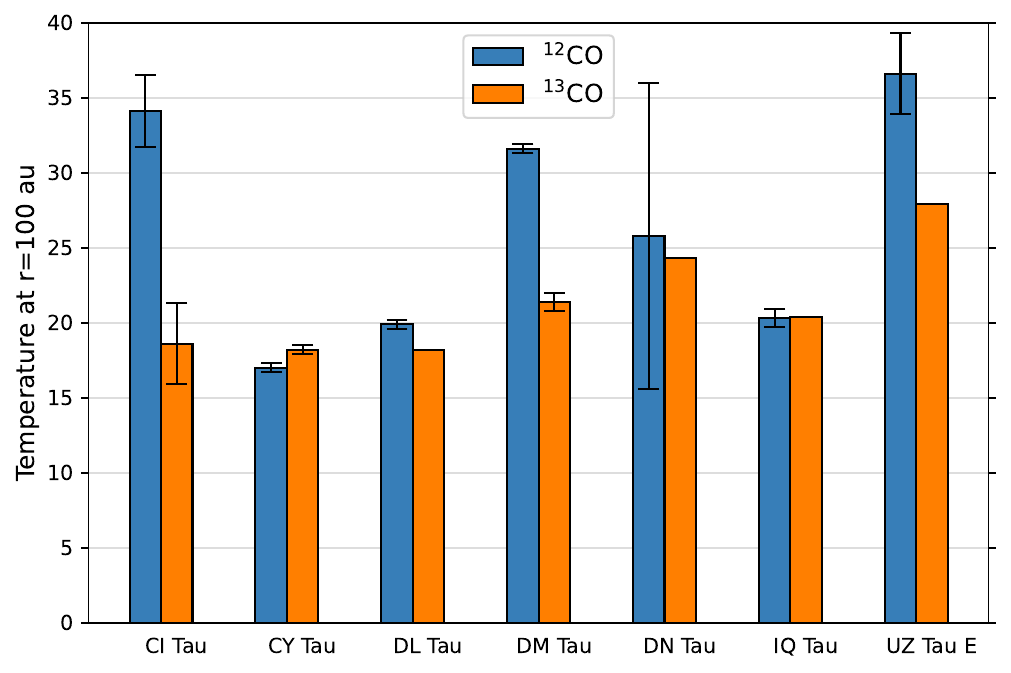}
\caption{Best-fit gas temperatures and their $1\sigma$ uncertainties for the observed $^{12}$CO and $^{13}$CO isotopologues at a disk radius of 100~au are shown. 
}
\label{fig:temp}
\end{figure}

The best-fit CO excitation temperature and column density profiles are presented in Table~\ref{tab:diskfit_phys} and compared in Fig.~\ref{fig:temp}. The (sub)millimeter CO  lines in disks are usually thermalized, and their excitation temperatures are thus close to the gas kinetic temperatures. As can be clearly seen, the derived CO temperatures in the outer disks at $r \sim 100$~au are low, $\sim 17-37$~K. Some of the derived temperatures at $r=100$~au are below the freeze-out temperature of CO of 20~K. \upd{From a theoretical perspective, temperatures below 20~K are expected in the outer, $\gtrsim 100-200$~au midplane regions in the disks around cool T~Tauri stars \citep[e.g.,][]{KH_87,Birnstiel_ea10a,Woitke_ea16}. Moreover, the presence of cold, $\lesssim 20$~K CO gas in the outer disks due to the grain evolution, nonthermal photo- or chemical desorption, or transport processes has been predicted by various detailed disk thermochemical models \citep[e.g.,][]{Semenov_ea06,Pietu_ea07,Krijt_ea20,Woitke_ea22,Gavino_ea23}. Alternatively, these low temperatures could be caused by the single power-law description of the disk temperatures in the DiskFit model, whereas in other disk models aiming at fitting the CO isotopologue or dust continuum emission, a temperature floor values have often been assumed to keep the disk midpane or CO layer temperatures above a certain thermal threshold value \citep[see, e.g.,][]{Andrews_ea09,Rosenfeld_ea13,Williams_Best14,Zhang_ea17_tw_hya,Schwarz_ea18,Tazzari_ea21b,Miotello_ea_PPVII_2022, Law_ea22a}.}

The best-fit temperature exponents $q$ are usually $\sim 0.4-0.7$. \upd{The $^{13}$CO and C$^{18}$O lines in CI~Tau are an exception, and are more consistent with flat temperature profiles.} These values are typical for outer disk regions around cool T~Tauri stars \citep[e.g.][]{Henning_Semenov13,Dutrey_ea14,Miotello_ea_PPVII_2022}. The less optically thin minor CO isotopologues probe a deeper, cooler part of the CO molecular layer, and thus show lower gas temperatures (depending on the vertical structure), see, e.g. \citet{Dutrey_ea17,Flores_ea2021}. The warmest CO gas is in the CI, DN, DM~Tau and UZ~Tau~E disks, with $T \sim 32-37$~K. The CI~Tau and DN~Tau have the highest stellar luminosities in our sample, $L_{*} \approx 0.8$ and $0.69 L_{\odot}$, respectively, and, hence, should have the warmest disk atmospheres (Table~\ref{tab:stellar_prop}). The disk of DM~Tau is flared and hence could be more efficiently heated by the central M~dwarf star, leading to a stronger temperature gradient \citep{Pietu_ea07}. \upd{The UZ~Tau~E disk may have a low gas mass of $\lesssim 0.003 M_{\odot}$ (see Appendix~\ref{append:sec:sample}), but this should not necessarily lead to a warmer CO molecular layer in this disk}. A more plausible explanation is that the luminosity of this binary system is higher than $\sim 0.35 L_{\odot}$ shown in Table~\ref{tab:stellar_prop}. Indeed, a higher luminosity value of $\sim 0.8-2.2 L_{\odot}$ have been adopted in other works \citep{Ricci_ea10,Guilloteau_ea16,Tripathi_ea17,Manara_ea_PPVII_2022}. 

\upd{Rather similar CO temperatures have been found for our disks in other studies. \citet{Law_ea22a} have constrained vertical structures of the CI~Tau and DM~Tau disks, using archival $\approx 0.1-0.4\arcsec$ ALMA data. They have found that the DM~Tau disk has an elevated CO emission surface, and inferred the $^{12}$CO temperatures between $\sim 20$~K in the outer disk and up to $\sim 35$~K in the inner disk at $r \lesssim 100$~au. These DM~Tau temperatures are also similar to those obtained in the earlier studies of \citet{Pietu_ea07} and \citet{Flaherty_ea20}. In \citet{Law_ea23a}, a detailed analysis of the vertical emission distribution and the gas temperatures using better $\approx 0.1-0.2\arcsec$ ALMA observations of the three main CO isotopologues have been performed for several disks, including DM~Tau. It has been found that the C$^{18}$O emission comes from the cold midplane, while the $^{12}$CO and $^{13}$CO emission comes from elevated heights, $z/r \sim 0.2-0.4$ (and the $^{12}$CO emitting surface is located higher than the $^{13}$CO surface). They derived the best-fit $^{12}$CO and $^{13}$CO temperatures at $r=100$~au of $38 \pm  0.5$ and $26 \pm  0.4$~K, respectively, which our DiskFit values of $31.6 \pm  0.1$ and $21.4 \pm  0.2$~K match rather well (see Table~\ref{tab:diskfit_phys}). The best-fit power-law exponents for the gas temperature profiles of $q \approx 0.42-0.56$ have been obtained by \citet{Law_ea23a}, which again are well matched by our DiskFit $q \approx 0.51-0.58$ values (within uncertainties).  
} 

\upd{In contrast to DM~Tau, the $^{12}$CO temperature at $r=100$~au in CI~Tau derived by \citet{Law_ea22a} is $\sim 48$~K, which is warmer than the best-fit  $^{12}$CO temperature of $\sim 34$~K derived in this study (see Table~\ref{tab:diskfit_phys}). One of the reasons for that is a lower spatial resolution of our $0.9\arcsec$ NOEMA CO data compared to the $\sim 0.1\arcsec$ ALMA data used by Law et~al., which makes it impossible for the DiskFit model to probe gas properties in the innermost $\sim 50-80$~au regions resolved with ALMA. Another reason is that the CO temperature profile constrained from the high-resolution ALMA data deviates from a simple power law in CI~Tau, with prominent gas temperature dips at the radii of 70 and 120~au, and a local gas temperature peak at 90~au (this peak coincides with a strong change in the CO surface emission profile and is close to a dust continuum emission ring), see \citet{Law_ea22a}. Our coarser NOEMA CO observations do not show these substructures, see Fig.~C.1. Hence, the DiskFit model that is not sensitive to these localized temperature deviations and is more strongly dominated by the CO data from the colder, $>100$~au region tends to underestimate the gas temperature in the inner disk.}

\begin{table*}
\caption{
\label{tab:diskfit_phys}
Best-fit gas thermal and CO column density profiles.}
\centering
\begin{tabular}{lllllllllllllll}
\hline\hline
Source & Line & Temperature at 100 au & Temperature exp. & Column density at 100 au & Column density exp. \\
       &      & (K)                   & $q$                    & (cm$^{-2}$) &  $p$              \\
\hline
CI~Tau & $^{12}$CO  (2-1)& $  34.1\pm    0.8$ & $  0.69\pm 0.01$ & $ 17.97$ & $  1.32$ \\ 
 & $^{13}$CO  (2-1)& $18.6\pm    0.9$ & $  0.05\pm 0.03$ & $ 16.18\pm   0.02$ & $  1.67\pm   0.11$  \\ 
 & C$^{18}$O (2-1) & [$18.7$] & $  0.09$ & $ 15.22\pm   0.01$ & $  1.59$ \\ 

\hline
CY~Tau & $^{12}$CO  (2-1)& $  17.0\pm    0.1$ & $  0.65\pm 0.01$ & $ 17.00$ & $  1.40$ \\ 
 & $^{13}$CO  (2-1)& $  18.2\pm    0.1$ & $  0.76\pm 0.01$ & $ 16.12\pm   0.06$ & $  1.75\pm   0.07$ \\ 
 & C$^{18}$O  (2-1)& [$  18.2 $]& [$  0.76$] & $ 14.88\pm   0.13$ & $  1.51\pm   0.09$ \\ 

\hline
DL~Tau & $^{12}$CO  (2-1)& $  19.9\pm    0.1$ & $  0.66\pm 0.01$ & $ 16.93$ & $  0.47$ \\ 
 & $^{13}$CO  (2-1)& $  18.2\pm    0.1$ & $  0.50\pm   0.01$ & $ 14.95\pm   0.05$ & $  0.18\pm   0.06$ \\ 
 & C$^{18}$O  (2-1)& [$  18.2$] & [$  0.50$] & $13.92\pm   0.08$ & [$  0.18$] \\ 

\hline
DM~Tau & $^{12}$CO  (2-1)& $  31.6 \pm 0.1$ & $  0.58\pm 0.05$ & $ 17.85\pm   0.39$ & $  2.75\pm   0.02$ \\ 
 & $^{13}$CO  (2-1)& $  21.4\pm    0.2$ & $  0.51\pm   0.01$ & $ 16.29\pm   0.02$ & $  1.57\pm   0.04$ \\ 
 & C$^{18}$O  (2-1)& [$  21.5$] & [$  0.5$] & $ 14.92\pm   0.02$ & $  0.53\pm   0.07$  \\ 

\hline
DN~Tau & $^{12}$CO  (2-1)& $  25.8\pm    3.4$ & $  0.62\pm 0.05$ & $ 16.36\pm   0.02$ & $  3.08\pm   0.08$ \\ 
 & $^{13}$CO  (2-1)& $  24.3\pm    0.1$ & $  0.65\pm   0.01$ & $ 15.52\pm   0.02$ & $  0.85\pm   0.05$  \\ 
 & C$^{18}$O  (2-1)& [$  24.3$] & [$  0.65$] & $ 14.64\pm   0.03$ & [$  0.85$] \\ 

\hline
IQ~Tau & $^{12}$CO  (2-1)& $  20.3\pm    0.2$ & $  0.41\pm 0.01$ & $ 16.85$ & $  1.40$  \\ 
 & $^{13}$CO  (2-1)& $  20.4\pm    0.1$ & $  0.37\pm   0.01$ & $ 14.96\pm   0.03$ & $  1.34\pm   0.59$ \\ 
 & C$^{18}$O  (2-1)& [$  20.4$] & [$  0.37$] & $ 14.04\pm   0.12$ & $  1.34\pm   0.59$ \\ 

\hline
UZ~Tau~E & $^{12}$CO  (2-1)& $  36.6\pm    0.9$ & $  0.67\pm 0.19$ & $ 16.65 \pm 0.02$ & $  1.28 \pm0.01$ \\ 
 & $^{13}$CO  (2-1)& $  27.9\pm    0.1$ & $  0.55\pm   0.01$ & $ 16.10 \pm   0.06$ & $  0.45\pm   0.04$ \\ 
 & C$^{18}$O  (2-1)& [$  27.9$] & [$  0.55$] & $ 15.34\pm   0.01$ & [$  0.50$] \\ 

\hline
\end{tabular}
\tablefoot{The parameters in square brackets have been kept fixed during the fitting. The best-fit temperatures are the excitation temperatures of the corresponding CO transitions. Please note that the best-fit column densities derived from the optically thick $^{12}$CO data can be underestimated.}
\end{table*}

\begin{figure}
\centering
\includegraphics[width=0.96\hsize,clip]{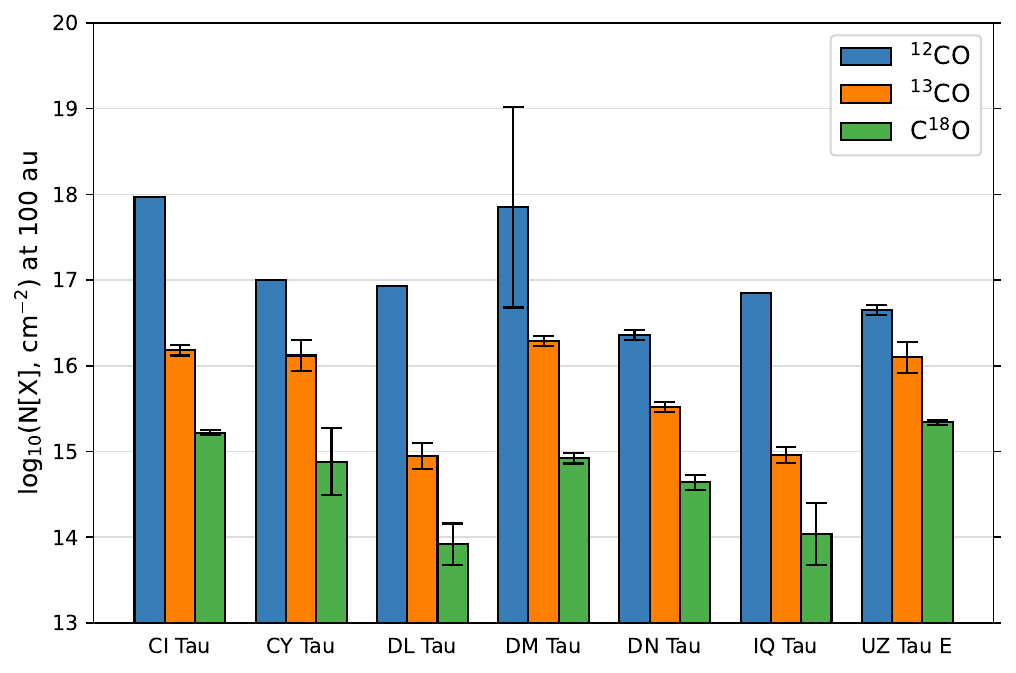}
\caption{Best-fit column densities and their $1\sigma$ uncertainties for the observed CO isotopologues at a disk radius of 100~au are shown.  The $^{12}$CO column densities can be underestimated due to the optical thickness of the $^{12}$CO emission. }
\label{fig:col_dens}
\end{figure}

The column density profiles at $r=100$~au are compared in Fig.~\ref{fig:col_dens} and are $\sim 2\times 10^{16}-10^{18}$~cm$^{-2}$ for $^{12}$CO, $\sim 10^{15}-2\times 10^{16}$~cm$^{-2}$ for $^{13}$CO, and $\sim 10^{14}- 2\times 10^{15}$~cm$^{-2}$ for C$^{18}$O, respectively. The power-law exponents of the column density profiles $p$ are $\sim 0.5-3.2$ for $^{12}$CO and vary between $0.2$ and $1.8$ for $^{13}$CO and C$^{18}$O. The $^{12}$CO~(2-1) emission is optically thick, and thus the best-fit $^{12}$CO column densities could be underestimated. \upd{(Constraints on the $^{12}$CO column density come in part from the optically thin line wings, and in part from the power law assumption).} The highest $^{12}$CO surface densities of $\sim 10^{18}$~cm$^{-2}$ were found in the CI~Tau and DM~Tau disks, whereas the lowest values of $\lesssim 10^{17}$~cm$^{-2}$ were found in the disks around CY, DL, DN,  IQ~Tau, and UZ~Tau~E. The column densities derived from the less optically thick $^{13}$CO~(2-1) are the highest for CI~Tau, CY~Tau, and DM~Tau ($\sim 2\times 10^{16}$~cm$^{-2}$), and the lowest in DL~Tau and IQ~Tau ($\lesssim 10^{15}$~cm$^{-2}$). For C$^{18}$O~(2-1), the highest column densities were obtained for CI~Tau, DM~Tau, and UZ~Tau~E ($\sim 10^{15}$~cm$^{-2}$), while the lowest values were found in DL~Tau and IQ~Tau ($\sim 10^{14}$~cm$^{-2}$). Thus, both the DL~Tau and IQ~Tau disks have the lowest CO surface densities in our sample. While for the IQ~Tau disk it can be explained by its low disk mass of $0.009 M_{\odot}$, the situation for DL~Tau is more interesting. The combined [CI] and CO observations of DL~Tau with ALMA and ACA by \citet{Sturm_ea22a} have revealed that carbon could be severely depleted by a factor of $\sim 50-150$ in the outer DL~Tau molecular layer. This is consistent with the apparently old age of the DL~Tau disk, $\sim 4-8$~Myr, as well as a measured high CO/dust size ratio, indicative of efficient radial drift and dust evolution, and associated chemical processing of CO \citep{Krijt_ea20,Powell_ea22,Sturm_ea22a}. Another estimate of the CO depletion for our sample has been derived for DM~Tau by \citet{McClure_ea16}, who found that the depletion is mild, within a factor of 5 or less.

\begin{figure}
\centering
\includegraphics[width=0.97\hsize,clip]{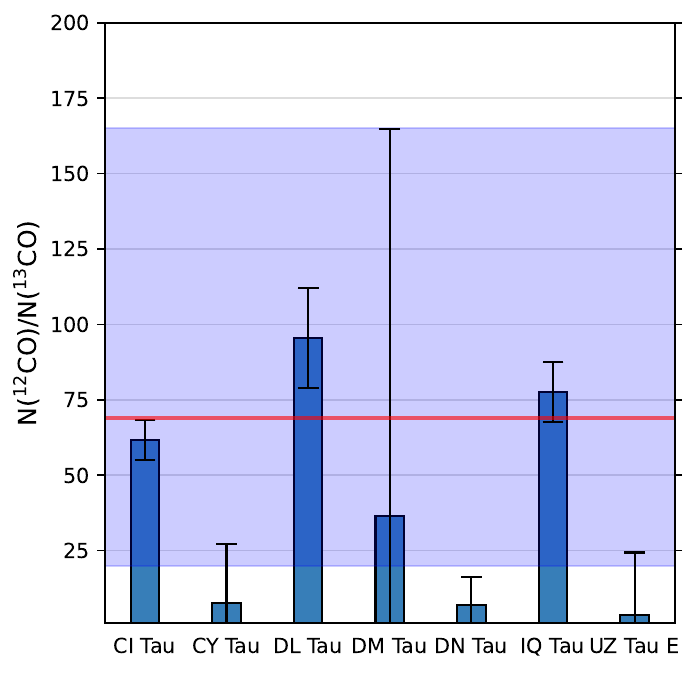}

\includegraphics[width=0.97\hsize,clip]{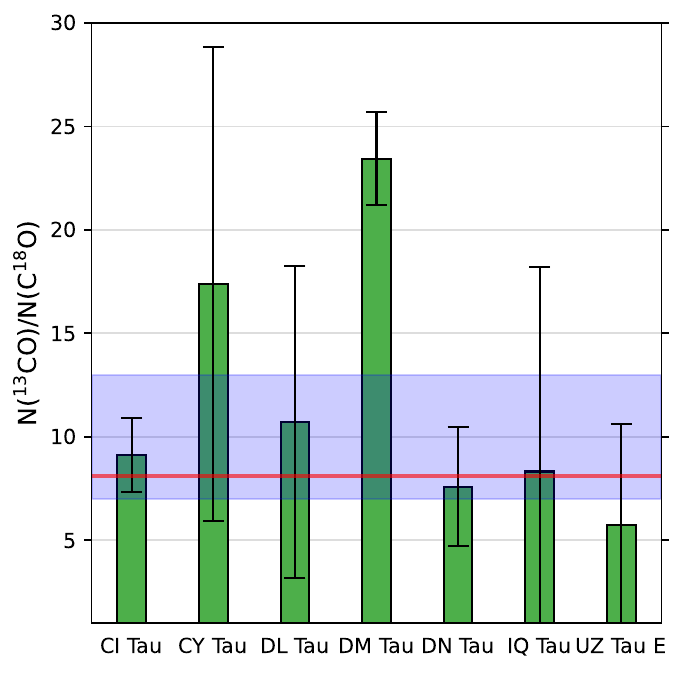}
\caption{Column density ratios of $^{12}$CO/$^{13}$CO (top) and $^{13}$CO/C$^{18}$O (bottom) at a disk radius of 100~au. The shaded area shows the isotopic ratios measured in disks or the ISM. The local ISM isotopic ratios are shown by the thick red line.}
\label{fig:iso_ratio}
\end{figure}

We used the best-fit column densities and calculated the corresponding isotopic ratios and the $1\sigma$ uncertainties for $^{12}$CO/$^{13}$CO and $^{13}$CO/C$^{18}$O (see Fig.~\ref{fig:iso_ratio}). Both the $^{12}$CO/$^{13}$CO and $^{13}$CO/C$^{18}$O ratios roughly fall within the ranges observed in disks and the ISM of $\sim 20-165$ and $\sim 8-12$, respectively \citep[e.g.][]{Pietu_ea07,Booth_ea19,Furuya_ea22,Yoshida_ea22}. The only exception is the $^{13}$CO/C$^{18}$O ratio in DM~Tau, where the C$^{18}$O column density could be underestimated due to the noisy C$^{18}$O~(2-1) data \upd{or, perhaps, could be the result of strong selective photodissociation of C$^{18}$O in the very extended, outer flared disk region} (see Fig.~\ref{fig:DM_Tau-C18O_comp}). 

\upd{Finally, we used the best-fit power-law profiles of the $^{13}$CO column densities to estimate disk gas masses. Since the best-fit $^{12}$CO column densities could be underestimated by DiskFit due to the optical thickness of the $^{12}$CO line, we assumed a range of $[20-69]$ for the $^{12}$CO/$^{13}$CO ratios and used it to recalculate the $^{12}$CO column densities. CO is heavily frozen-out in the midplane of the cold, $>100$~au disk regions probed by our NOEMA observations, hence we have adopted a non-ISM $^{12}$CO/H$_2$ ratio of $\sim 10^{-6}$  to constrain the underlying gas surface density profile \citep[e.g.,][]{SW2011,Miotello_ea18a,Krijt_ea20,Trapman_ea22}. By taking all these assumptions into account, the resulting disk gas masses vary between $\sim 10^{-3} - 0.2~M_\odot$ with considerable uncertainties, see Table~\ref{tab: disk_gas_masses}. Compared to the rescaled dust masses adopted from the previous studies (see Table~\ref{tab:disk_prop}), the gas masses calculated from our $^{13}$CO data are somewhat higher by a factor of $\sim 2-3$, except for the disks around DL~Tau, DN~Tau and IQ~Tau, which have lower CO-derived masses compared to their rescaled dust-derived masses. This is not surprising because the DL~Tau, DN~Tau, and IQ~Tau disks have low C$^{18}$O~(2-1) fluxes and low best-fit C$^{18}$O column densities, indicating a low amount of CO gas. The CO-derived mass of the extended, $R_{\rm CO} \sim 620$~au disk around DL~Tau deviates the most from its dust mass, once again suggesting that this peculiar system may have more severe CO depletion than in other disks \citep[see also][]{Sturm_ea22a}, while the compact disks around DN~Tau and IQ~Tau have low total gas masses. The most massive disk in our sample is DM~Tau with a CO-derived mass of $\sim 0.06-0.2M_\odot$, which is somewhat higher than its mass of $\lesssim 0.045M_\odot$ derived from the {\em Herschel} HD observations \citep{McClure_ea16,Trapman:2017tn}. A more detailed study of the gas densities and masses in PRODIGE sources derived by the physical-chemical modeling will be presented in \citet{Franceschi_ea24a}. 
}

\begin{table}
\caption{Disk gas masses derived from the $^{13}$CO~(2-1) column densities.}
\label{tab: disk_gas_masses}
\centering
\begin{tabular}{lccc}
\hline\hline
Source &  $^{12}$CO/$^{13}$CO = 20  &  $^{12}$CO/$^{13}$CO = 69  & $M^{\rm dust}_{\rm disk}\times 100$\\
\cline{2-4}
       &  ($M_\odot$) & ($M_\odot$) & ($M_\odot$)\\
       
\hline
CI~Tau &  $0.04$  & $0.15$   & 0.016\\
CY~Tau &  $0.04$  & $0.12$   & 0.019\\
DL~Tau &  $0.005$ &  $0.02$  & 0.036\\
DM~Tau &  $0.06$  &  $0.2$   & 0.025\\
DN~Tau &  $0.003$  &  $0.01$ & 0.013\\
IQ~Tau &   $0.001$ &  $0.004$& 0.009\\
UZ~Tau~E & $0.01$ & $0.04$   & 0.018\\
\hline
\end{tabular}
\tablefoot{Columns 2 and 3: Disk masses derived assuming two different $^{12}$CO/$^{13}$CO ratios. Column 4: Disk masses derived from the dust continuum that were taken from Table~\ref{tab:disk_prop}.}
\end{table}

\subsection{Correlations between disk physical and chemical properties}
\label{subsec:statistics}

In this subsection, we \upd{discuss trends among the} various disk physical and chemical parameters. Given a modest size of our disk sample with $<10$ systems and a biased selection procedure, the results of such analysis should be taken with care. We compared the dust continuum fluxes, CO emission radii, CO temperatures, disk masses derived from dust masses rescaled by an assumed gas/dust ratio of 100, and stellar masses versus the absolute and normalized integrated and peak fluxes, CO column densities, ages, etc., and found only a few potential correlations. For each correlation, we calculated the corresponding coefficient of determination, which is equal to the Pearson coefficient $r$ squared, by taking the data uncertainties when necessary.

\begin{figure}[]
\centering
\includegraphics[width=0.7\hsize,angle=270]{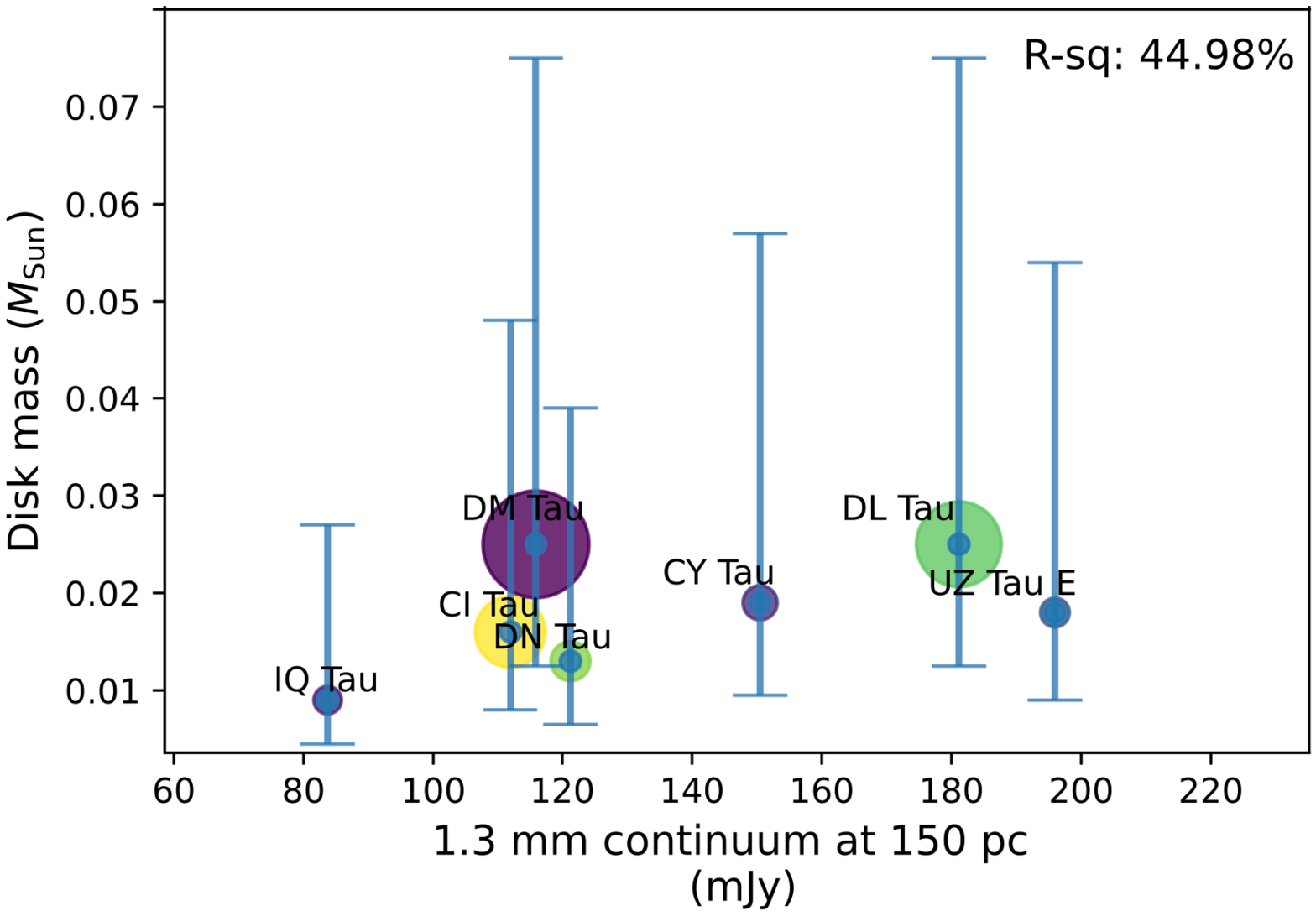}
\caption{\label{fig:cont1mm_m_disk}
Correlation between the thermal dust continuum fluxes at $1.3$~mm and the disk gas masses \upd{derived by rescaling disk dust masses}. The size of the disk symbol reflects the disk size, and the color of the symbol designates the luminosity of the central star (darker colors correspond to lower luminosities). The coefficient of determination (corresponding to the Pearson coefficient $r$ to the power of two) was calculated by taking the data uncertainties into account and is shown in the top right corner.}
\end{figure}

First, we compare the 1.3~mm continuum and the \upd{rescaled disk dust} masses in Fig.~\ref{fig:cont1mm_m_disk}. As can be clearly seen, \upd{despite the fact that dust masses are derived from continuum emission, there is no strong correlation between these quantities due to significant uncertainties in the derived dust masses and a limited disk sample size. A similar absence of a clear correlation was found between the CO-derived gas masses and the dust continuum, again due to gas mass uncertainties.} 

\begin{figure*}[]
\centering
\includegraphics[width=0.45\hsize]{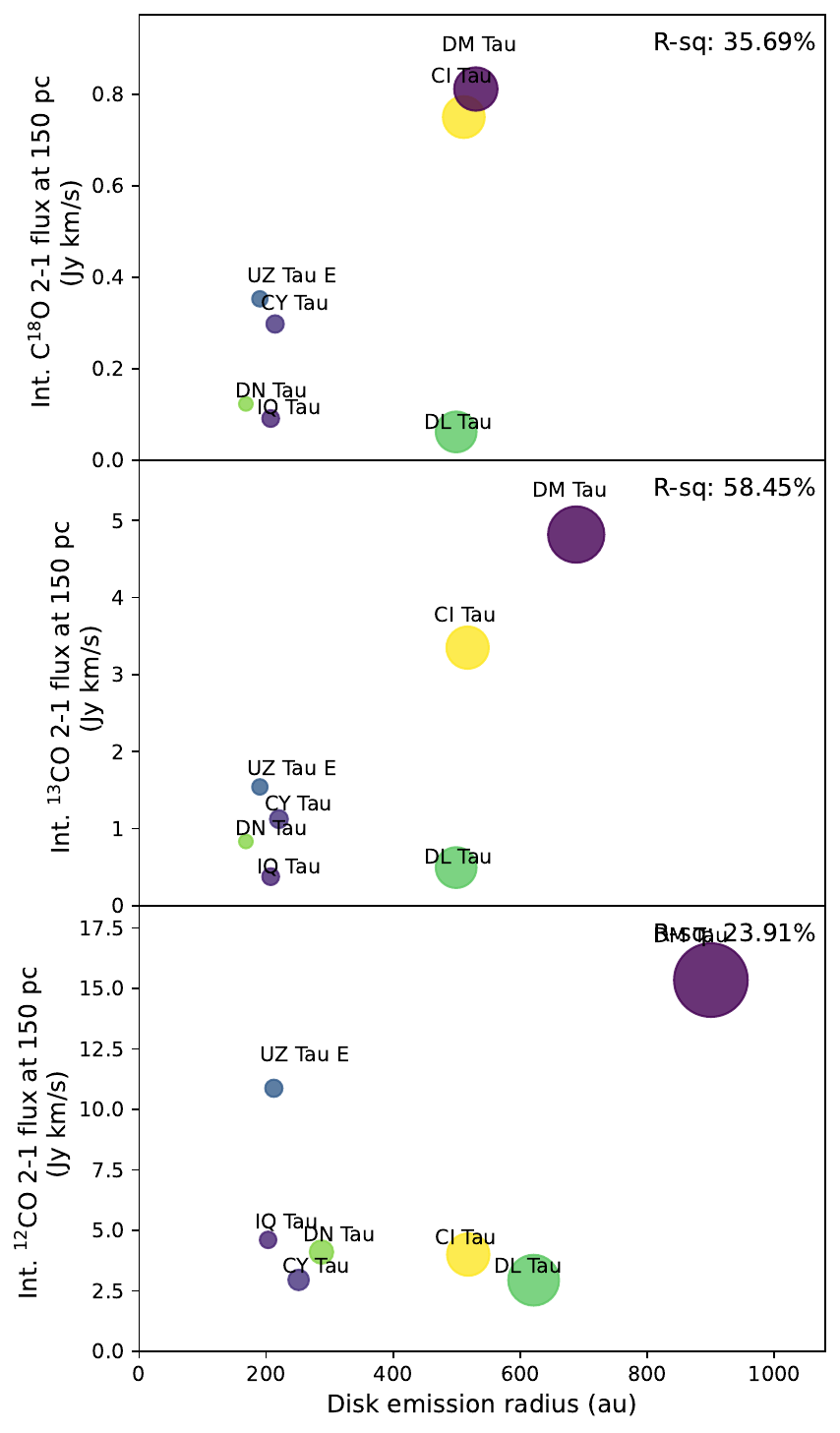}
\includegraphics[width=0.45\hsize]{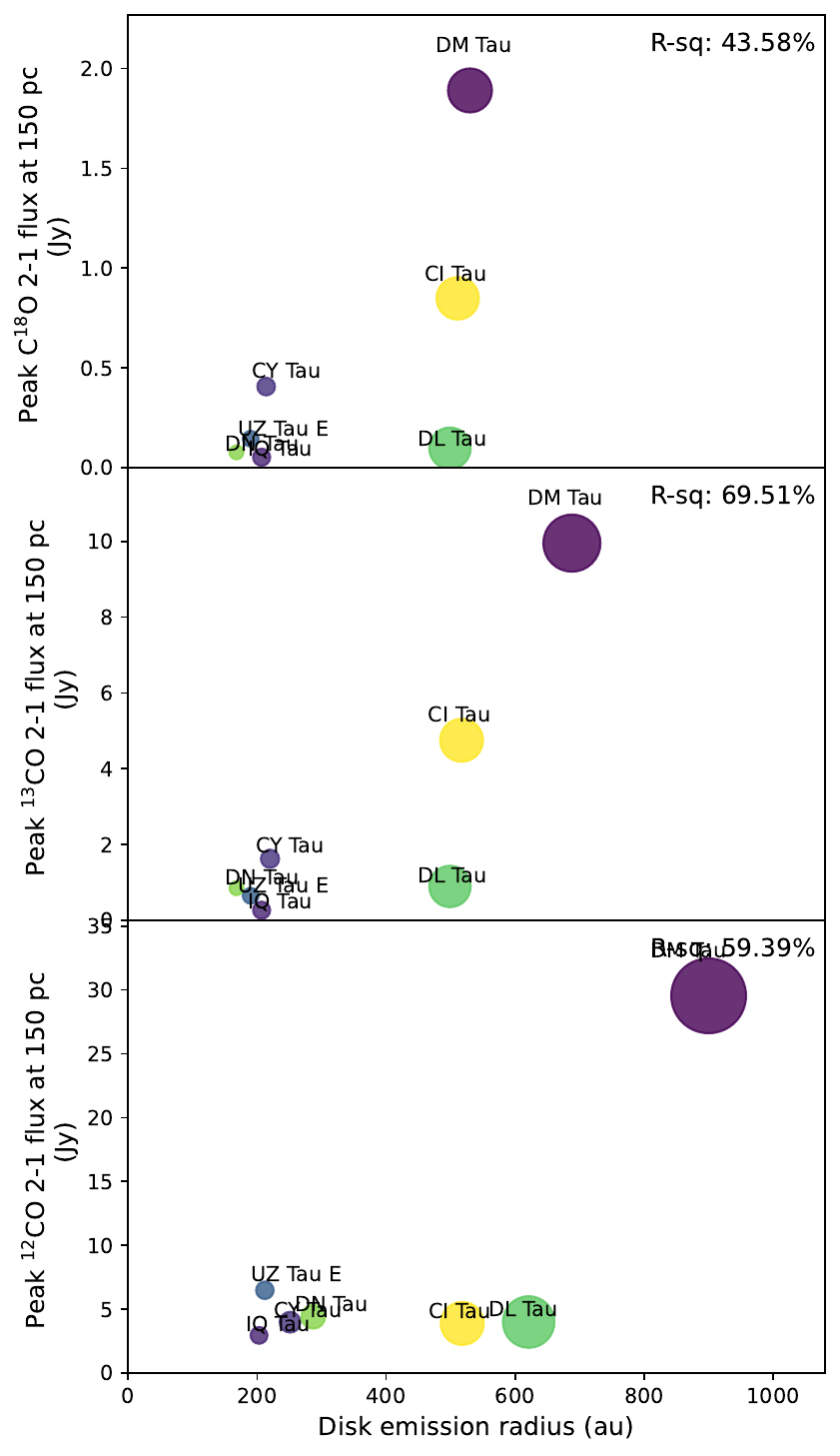}
\caption{\label{fig:radius_flux}
Correlation between the CO gas radii and the integrated (left) and peak (right) fluxes. The $^{12}$CO-, $^{13}$CO-, and C$^{18}$O-based data are shown from bottom to top. The size of the disk symbol reflects the disk emission size for the given CO isotopologue, and the color of the symbol designates the luminosity of the central star (darker colors correspond to lower luminosities). The coefficients of determination (corresponding to the Pearson coefficients $r$ to the power of two)  were calculated by taking the data uncertainties into account and are shown in the top right corner of each subplot.}
\end{figure*}

Next, we compare the best-fit CO emission radii with the corresponding integrated and peak fluxes in Fig.~\ref{fig:radius_flux}. \upd{While the optically thick $^{12}$CO~(2-1) data show a potential correlation between the peak flux $S_{\nu}$ and the disk gas radius $R_\mathrm{CO}$ (bottom right panel), for the optically thin CO isotopologues there is a dependence between both the peak and integrated fluxes, and the disk gas radius. In all these cases, the DL~Tau disk with its very low CO fluxes (wrt its large, $\sim 600$~au CO radius) appears as an outlier. The same trend is visible for the CO fluxes normalized to the dust continuum vs $R_\mathrm{CO}$. The likely interpretation of this potential trend is a combination of similar cold temperatures in the outer regions of our T~Tau disk sample, coupled with similar radial CO density profiles off-set by distinct (but yet not too different) absolute CO column densities.} 

Interestingly, the correlation between \upd{the disk radius $R$ and the disk luminosity/flux $F$, $R \propto F^{0.5}$}, has been observationally inferred for the dust continuum for many disks in Chamaeleon, Lupus, Ophiuchus, and Taurus, and for the $^{12}$CO emission in the Lupus and Upper Scorpius disks \citep[and references therein]{Manara_ea_PPVII_2022}. \upd{Such a correlation should in general hold when 1) emission is optically thick and 2) the temperature spread is relatively narrow.} Manara et al. have mentioned several explanations for such correlation, such as (1) dust grain growth and radial drift in a low viscosity regime or when slowed by pressure bumps in the disk substructures, or (2) the presence of optically thick dust substructures. \upd{In the recent study by \citet{Zagaria_ea23a}, semi-analytic and detailed DALI thermo-chemical modeling of the disk CO fluxes have been performed, assuming a viscous accretion model and an MHD-wind-based accretion model. They have found that indeed the CO fluxes correlate with the disk radii, which can be constrained from the fluxes even when the disks are not spatially resolved. They have also found that the disk models are able to reproduce the CO fluxes observed in the Lupus star-forming region only when a CO depletion factor of $\sim 40-100$ is introduced in the model atop of the ISM abundance $x_\mathrm{CO} = 10^{-4}$, which is representative of severe CO depletion and photodissociation in the outer disk regions.}

\begin{figure}[]
\includegraphics[width=0.95\hsize]{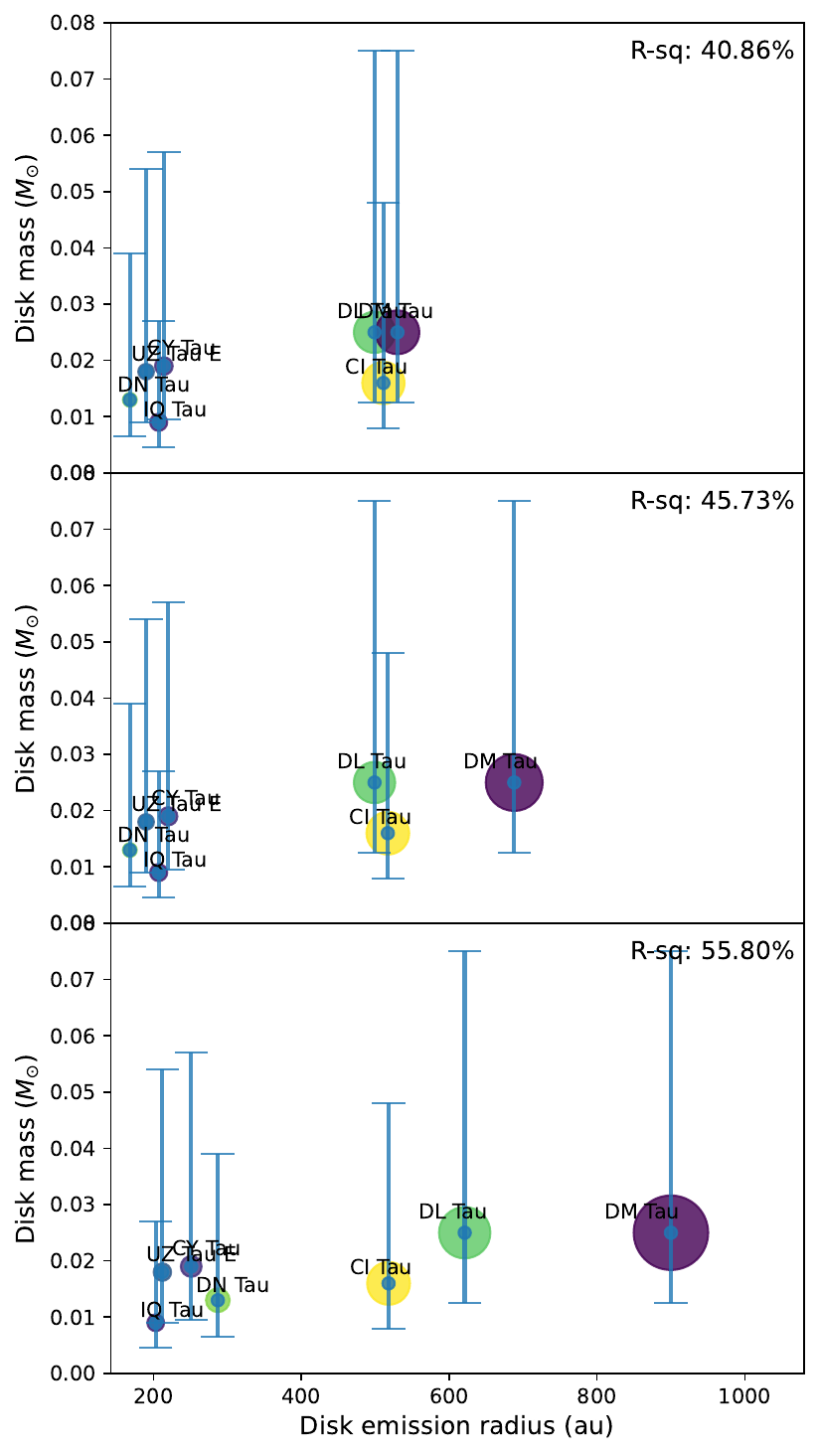}
\caption{\label{fig:radius_others}
Same as Fig.~\ref{fig:radius_flux} but for the correlation between the CO gas radii and the disk masses. 
}
\end{figure}

Another quantity that could correlate with the \upd{CO isotopologue emission radius in our sample is the dust-derived disk mass (see Fig.~\ref{fig:radius_others}). A naive explanaition for such a trend could be again the radial dust drift and slower depletion of the mm dust reservoirs in larger disks compared to the compact ones. Better statistics regarding disk gas properties from larger, $\gtrsim 30-50$ disks' observational surveys is needed to verify or falsify our findings.}

\section{Summary and conclusions}
\label{sec:conc}
\upd{We presented the large MPG-IRAM Observatory Program PRODIGE at NOEMA (L19ME; PIs: P.~Caselli and Th.~Henning), which is aimed at characterizing the properties of eight Class~II T~Tauri disks in Taurus via a 1~mm line survey (2020-2024). The chosen CI~Tau, CY~Tau, DG~Tau, DL~Tau, DM~Tau, DN~Tau, IQ~Tau, and UZ~Tau~E systems include both compact and extended disks without severe substructures. To deal with a large volume of interferometric data in a coherent manner, we developed the automatic Gildas/Imager pipeline for line selection, imaging, self-calibration, Kepler deprojection, and the S/N boosting. For this first paper, the combined $^{12}$CO, $^{13}$CO, and C$^{18}$O~$J=2-1$ data taken at the $0.9\arcsec$ and 0.3~km\,s$^{-1}$ resolution with the $8-12$~mJy/beam sensitivity were analyzed and modeled.} 

\upd{We detected the emission of the three main CO isotopologues in all the disks (though C$^{18}$O~(2-1) was only marginally detected in IQ Tau). The foreground cloud absorption affects the optically thick $^{12}$CO~(2-1) emission in CI~Tau, CY~Tau, DG~Tau, DL~Tau, and UZ~Tau~E, and also in $^{13}$CO in DL~Tau. We spatially resolved the CO emission in all of the disks, and inferred their outer gas radii, ranging from $\sim 200$ to 900~au. The peak line brightness temperatures are $\sim 7-16$~K for $^{12}$CO~(2-1), $1-8$~K for $^{13}$CO~(2-1), and $0.2-2$~K for C$^{18}$O~(2-1), respectively, indicating that the outer disk gas is cold and that the $^{13}$CO~(2-1) and C$^{18}$O~(2-1) lines are mostly optically thin. } 

\upd{We found that the CO~(2-1) fluxes may scale with the disk outer radius (except that of DL~Tau). This may suggest that the assumed radial distribution of the gas is similar in all of our disks, which differ only by the normalization (a total gas mass). The DL~Tau system stands out with a large CO radius but faint $^{13}$CO and C$^{18}$O emission. Homogeneous depletion of CO cannot account for this discrepancy and its gas radial distribution should differ from those in the other targeted disks. The UZ~Tau~E disk shows evidence of non-Keplerian motions, most likely due to the tidal interactions with its UZ~Tau~W binary companion located $\sim 600$~au away.}

\upd{We fit each line in the Fourier space using the power-law DiskFit model. 
The discerned gas temperatures at 100~au are low, $\sim 17-37$~K, and close or slightly above the CO evaporation temperature. 
The temperatures of the CO molecular layers decline radially with a power-law exponent of $ \sim 0.5-0.7$.  When strong enough, the $^{13}$CO~(2-1) emission is indicative of a flatter temperature profile with an exponent of $\sim 0 - 0.4$, and it usually has lower temperatures than the temperatures probed by the $^{12}$CO line
in the extended disks, consistent with the vertical temperature gradients. The difference between the $^{12}$CO and $^{13}$CO temperatures is not systematic though.}

\upd{We found that the $^{12}$CO, $^{13}$CO, and C$^{18}$O column densities at $r=100$~au are $\sim 10^{17} - 10^{18}$~cm$^{-2}$, $\sim 10^{15} - 10^{16}$~cm$^{-2}$, and $\lesssim 10^{14} - 10^{15}$~cm$^{-2}$, respectively. The lowest $^{13}$CO and C$^{18}$O column densities were found in the DL~Tau and IQ~Tau disks. Most disks have a $^{13}$CO column density exponent $p$ of about $1.5$, with a notable exception for DL~Tau, where it is almost zero.
The inferred CO column densities point to CO depletion in the outer disk regions. In addition, there could be selective depletion of carbon inside the inner, $\lesssim 150-200$~au, region in the DL~Tau disk, making the overall CO budget of this extended $\sim 600$~au sized disk much lower than in the other sources. These results support the previous finding based on ALMA observations that CO can be strongly depleted in the DL~Tau disk by a factor of $\gtrsim 100-200$ (atop of the depletion due to CO photodissociation and freeze-out).}

\upd{The best-fit isotopic $^{12}$CO/$^{13}$CO ratios of $\sim 5-80$ and the $^{13}$CO/C$^{18}$O ratios of $\sim 6-23$ broadly agree within the uncertainties with the previously discerned isotopic values in disks and the local ISM of $\sim 20-165$ and $\sim 8-12$, respectively. For CI~Tau, CY~Tau, DM~Tau, and UZ~Tau~E, the disk gas masses estimated from the best-fit $^{13}$CO radial profiles and the disk gas radii are $\sim 0.001 - 0.2~M_\odot$ (with large uncertainties), and they are a factor of two to three higher than the rescaled dust disk masses. 
}

\upd{Finally, we investigated the dependences between various distinct disk and stellar parameters. Owing to the limited size of our sample, the only clear trends that we could identify are between the CO emission sizes and the CO fluxes or disk masses. Similar relations have been found for the $^{12}$CO emission observed in the Lupus and Upper Scorpius disks, as well as for the dust continuum observations of disks in the Chamaeleon, Lupus, Ophiuchus, and Taurus star-forming regions. }

\begin{acknowledgements}
This work is based on observations carried out under project numbers L19ME and S19AW (for DN\,Tau and CI\,Tau) with the IRAM NOEMA Interferometer. IRAM is supported by INSU/CNRS (France), MPG (Germany) and IGN (Spain). D.~S., Th.~H., G.~S.-P., R.~F, K.~S., and S.~vT acknowledge support from the European Research Council under the Horizon 2020 Framework Program via the ERC Advanced Grant No. 832428-Origins. \upd{A.~F. acknowledges funding from the European Research Council via the ERC advanced grant SUL4LIFE (GAP-101096293).  A.~F. also thanks the Spanish MICIN for funding from PID2019-106235GB-I00.}
 This work was partly supported by the Programme National “Physique et Chimie du Milieu Interstellaire” (PCMI) of CNRS/INSU with INC/INP co-funded by CEA and CNES. This research made use of NASA's Astrophysics Data System. Some of the figures in this paper were made with the \textsc{matplotlib} package \citep{Hunter2007}.
\end{acknowledgements}


\begin{appendix}
\section{Additional information about the disk sample}
\label{append:sec:sample}
\paragraph{\it CI~Tau:} The CI~Tau system consists of a single star surrounded by a $\sim 175$~au dust disk and a $\sim ~520$~au CO gas disk \citep{Long_ea18a,Long_ea19_taurus,Law_ea22a}. The stellar dynamical mass is $0.9M_\odot$, while other mass estimates derived from the stellar models are $0.8-1.29M_\odot$ \citep{Ricci_ea10,Simon_ea19,Gangi_ea22}. A hot Jupiter planet with a 9-day orbital period has been found in this system \citep{Johns_Krull_ea16,Flagg_ea19}, making it a (pre-)transitional system. The high-resolution ALMA observations by \citet{Clarke_ea18,Konishi_ea18,Long_ea18a,Rosotti_ea21} have revealed several emission rings and gaps in the dust continuum. The dust gaps with widths of $\sim 9-22$ are located at radii of $\sim 12-14$, $39-48$ and $100-120$~au, suggesting that there could be up to four planets interacting with the disk. There is a hint for the presence of a $^{13}$CO gap at $\sim 60$~au \citep{Rosotti_ea21}. The disk mass derived from dust emission varies between $\sim 0.015$ and $0.05$ solar masses \citep{Andrews_Williams07,Ricci_ea10,Guilloteau_ea16,Ribas_ea20}. 

\paragraph{\it CY~Tau:}
CY~Tau is the lowest mass star in our sample, with a dynamical mass of only $0.3M_\odot$ \citep{Simon_ea19}. Its mass estimates derived by various stellar models deviate considerably from the dynamical mass, and are $\sim 0.35-0.5M_\odot$ \citep{Ricci_ea10,Simon_ea19,Pegues_ea21_masses}. The single CY~Tau star is surrounded by a prominent disk with a dust radius of $\sim 100$~au and a gas radius of $\sim 200-300$~au \citep{Guilloteau_ea14,Herczeg_Hillenbrand14,Guilloteau_ea16,Powell_ea19}. The disk mass derived from dust emission is $\sim 0.01-0.077M_\odot$ \citep{Ricci_ea10,Guilloteau_ea11a,Guilloteau_ea16,Ribas_ea20}. The disk gas mass derived from the fitting the dust lane locations and the disk surface density is $0.06M_\odot$ \citep{Powell_ea19}.

\paragraph{\it DG~Tau:}
 \textsg{DG~Tau is a young, $\sim 0.3-0.5$~Myr Class I/II object,}
partly enshrouded by an extended envelope.  It shows both thermal and nonthermal radio emission associated with a bipolar jet, an outflow, and shocks \citep{Guilloteau_ea12a,Purser_ea18,Garufi_ea22}. The stellar mass of the targeted DG~Tau star is about $0.8M_\odot$ \citep{Banzatti_ea19a}. The DG~Tau disk has a CO gas radius of $\sim 200-300$~au \citep{Guilloteau_ea11a}. The disk mass estimates vary wildly, $\sim 0.03-0.4M_\odot$ \citep{Isella_ea09,Guilloteau_ea11a,Ribas_ea20}.

\paragraph{\it DL~Tau:}
DL~Tau is the most massive star in our sample, with a dynamical mass $M_* \approx 1\,M_\odot$ (non-dynamical mass estimates vary between 0.8 and $1.1M_{\odot}$), see \citet{Ricci_ea10,Long_ea18a,Simon_ea19,Gangi_ea22}. The DL~Tau system has a dust disk with a radius of $\sim 150$~au and a gas disk with a radius of $~\sim 500$~au \citep{Long_ea18a,Long_ea22a}. The $0.1\arcsec$ ALMA observations by \citet{Long_ea18a} and \citet{Long_ea20} have revealed three dust gaps with widths of $\sim 7-26$~au, located at $\sim 40$, $67$, and $89$~au \citep{Long_ea18a}. Fainter substructures at $\gtrsim 150$~au have been found in the dust emission \citep{Jennings_ea22_gaps}. The derived disk mass from dust observations is $\sim 0.01-0.1M_\odot$ \citep{Andrews_Williams07,Ricci_ea10,Guilloteau_ea11a}. The age of this system is poorly estimated. While \citet{Ricci_ea10} have derived an age of $\sim 2.2$~Myr, which is in agreement with other estimates, \citet{McClure_ea19} has estimated that the DL~Tau system could be much older, $\sim 7.8$~Myr.

\paragraph{\it DM~Tau:}
The well-studied DM~Tau system is a single star with a dynamical mass of $0.55M_{\odot}$ surrounded by a large $\sim 200$~au dust and a $\sim 600-800$~au gas disk \citep{Guilloteau_ea11a,Simon_ea19,Long_ea20}. The age of DM~Tau is $\sim 1-5$~Myr \citep{Guilloteau_ea11a}. The DM~Tau disk shows two prominent gaps in dust emission at $\sim 3$ and $21$~au, with more shallow gaps at $\sim 70$, $83$ and $\sim 112$~au \citep{Kudo_2018,Hashimoto_ea21,Francis_ea22}. The DM~Tau disk is turbulent, with discerned $\alpha \sim 10^{-2}-8\times10^{-2}$ \citep{Guilloteau_ea12a,Flaherty_ea20,Francis_ea22}. The disk gas mass derived from the modeling HD~(1-0) emission is $\sim 0.004-0.045M_{\odot}$, depending on the assumed disk thermal structure \citep{McClure_ea16,Trapman:2017tn}.

\paragraph{\it DN~Tau:}
DN~Tau has a dynamical mass of $0.87M_{\odot}$ and is surrounded by a disk with a $\sim 100$~au dust and a $\sim 240$~au gas radius \citep{Ricci_ea10,Long_ea18a,Simon_ea19}. The dust emission shows the presence of a $\sim 23$~au wide gap at $\sim 60$~au \citep{Long_ea18a}. The disk mass derived from dust continuum data is $\sim 0.01-0.04M_{\odot}$ \citep{Guilloteau_ea16,Ribas_ea20}.

\paragraph{\it IQ~Tau:}
IQ~Tau has a dynamical mass of $0.74M_{\odot}$ and is surrounded by a disk with a $\sim 100$~au dust and a $\sim 220$~au gas radius \citep{Ricci_ea10,Long_ea18a,Simon_ea19}. The non-dynamical mass estimates span a broad range, $0.4-0.7M_{\odot}$. The IQ~Tau disk shows the presence of a $\sim 7$~au dust continuum gap at $\sim 40$~au \citep{Long_ea18a}. The disk mass derived from dust continuum data is $\sim 0.007-0.02M_{\odot}$ \citep{Ricci_ea10,Guilloteau_ea16,Ribas_ea20}.

\paragraph{\it UZ~Tau~E:}
UZ~Tau~E is a spectroscopic binary EXOr system with a separation of $\sim 0.03$~au, which is gravitationally linked with another close binary UZ~Tau~W system at a $3.6\arcsec$ separation \citep{Hales_ea20,Long_ea22a}. 
The combined mass of the UZ~Tau~E is $1.2M_{\odot}$. The UZ~Tau~E is surrounded by a disk with a $\sim 80$~au dust and a $\sim 200-300$~au gas radius \citep{Ricci_ea10,Long_ea22a}. It has a small inner cavity with a radius of $\lesssim 10$~au and a shallow, $~\sim 7$~au gap at the radius of $\sim 70$~au \citep{Long_ea18a,Jennings_ea22_gaps}. The dust disk mass is estimated to be $\sim 1-2.2\times 10^{-4}M_{\odot}$, while the gas mass derived from the fitting optically thin CO data is $\sim 3.2\times10^{-3}M_{\odot}$ \citep{Ricci_ea10,Guilloteau_ea16,Long_ea18a,Hales_ea20}. The high, $\sim 0.03-0.07$ dust/gas ratio in the UZ~Tau~E disk is similar to the values found in other EXors as well as Lupus disks \citep{Hales_ea20}. 

\section{Spectral setup 1 (2019-2020)}
\label{append:sec:setup1}

\begin{figure*}[!ht]
\centering
\includegraphics[angle=90,width=0.87\hsize, trim = 2cm 2cm 2cm 2cm, clip]{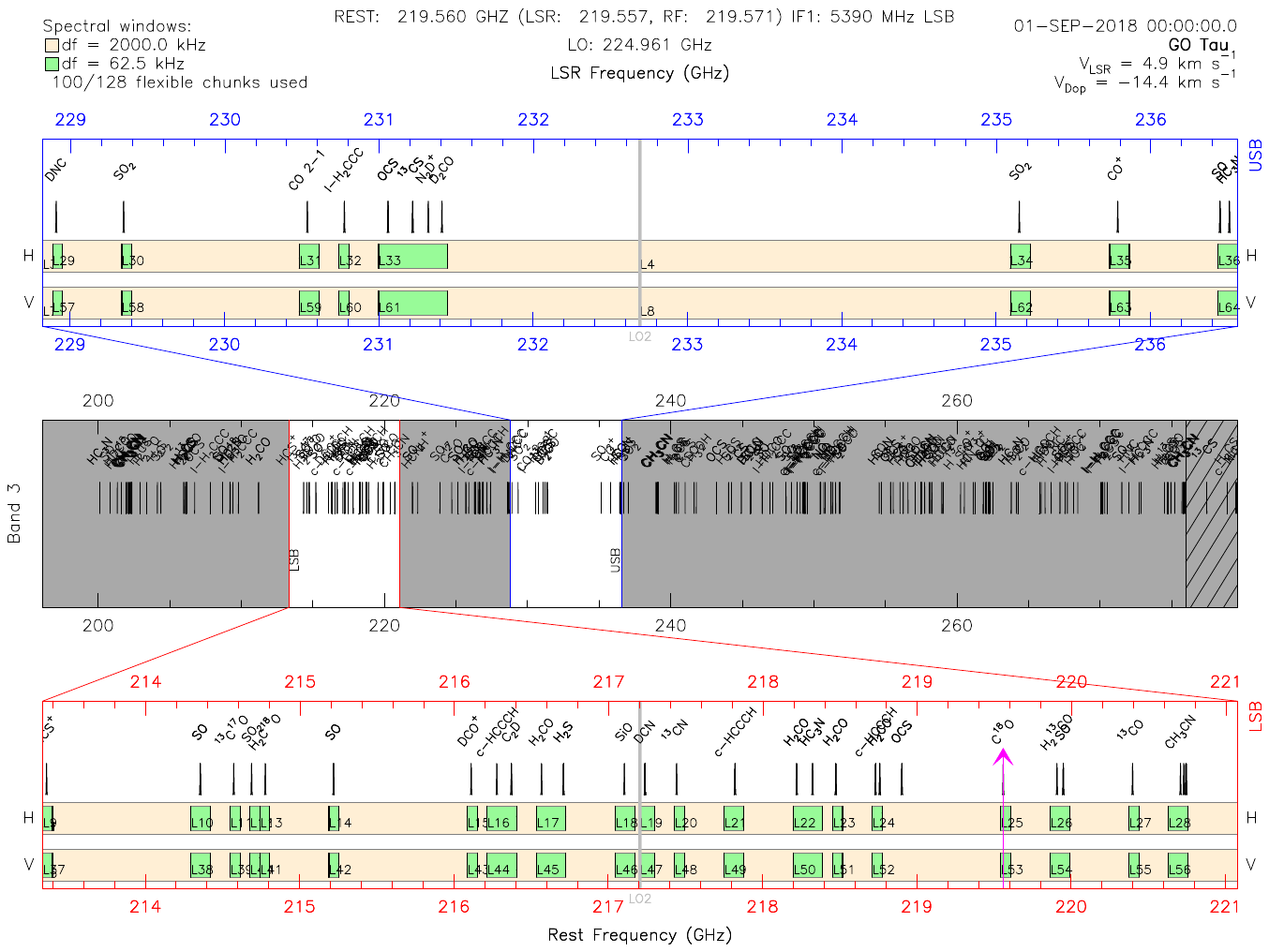}
\caption{\label{fig:setup1}
First NOEMA/PolyFiX setup with an LO frequency of $\approx 224.9$~GHz that has been observed during 2020. The following key disk molecular species have been targeted at the native 62.5~kHz ($\sim 0.1$~km\,s$^{-1}$) spectral resolution: $^{12}$CO, $^{13}$CO, C$^{18}$O, p-H$_2$CO, DCO$^+$, DCN, DNC, c-C$_3$H$_2$, HC$_3$N, N$_2$D$^+$, etc.}
\end{figure*}

\begin{table*}
\caption{Spectroscopic line parameters for the PolyFiX setup~1 observations (2019-2020).}
\label{tab:line_dat_s1}
\centering
\begin{tabular}{llllll}
\hline\hline
Frequency   &     Molecule      & Transition         & $E_{u}$   & $S_{ij}\mu^{2}$  & Catalog \\
(GHz)       &                   &                    & (K)       & (D$^{2}$)        &         \\
\hline
213.360650  & HCS$^+$           &  5-4               & 30.72     &  19.2            & CDMS  \\
214.357039  & SO $3\Sigma$ $\nu = 0$  & 7(8)-7(7)      & 81.24     &  0.44            & CDMS  \\
214.573873  & $^{13}$C$^{17}$O  & J=2-1, F=5/2-5/2   & 15.45     &  0.05            & CDMS  \\
214.778436  & H$_2$C$^{18}$O    & 3(1,2)-2(1,1)      & 32.48     & 43.50            & SLAIM \\
215.220653  & SO $3\Sigma$ $\nu = 0$  & 5(5)-4(4)      & 44.10     & 11.31            & CDMS  \\
216.112582  & DCO$^+$           & 3-2                & 20.74     &142.00            & CDMS  \\
216.278756  & c-HCCCH           &  3(3,0)-2(2,1)     & 19.47     & 45.62            & CDMS  \\
216.372830  & C$_2$D $^*$& N=3-2, J=7/2-5/2, F=9/2-7/2   & 20.77     &  2.53            & CDMS  \\
216.373320  & C$_2$D $^*$& N=3-2, J=7/2-5/2, F=5/2-3/2   & 20.77     &  1.87            & CDMS  \\
216.568651  & H$_2$CO           & 9(1,8)-9(1,9)      &174.00     &  3.48            & JPL   \\
216.710436  & H$_2$S            & 2(2,0)-2(1,1)      & 84.00     &  2.06            & CDMS  \\ 
217.104938  & SiO               & 5-4                & 31.25     & 47.99            & CDMS  \\
217.238400  & DCN               & J=3-2, F=2-1       & 20.85     & 18.11            & JPL   \\
217.238631  & DCN               & J=3-2, F=2-2       & 20.85     & 38.81            & JPL   \\
217.443722  & $^{13}$CN $^*$ & N=2-1, J=5/2-3/2, F1= 2-1, F=1-1 & 15.68 & 1.09           & CDMS  \\
217.822148  & c-HCCCH           & 6(1,6)-5(0,5)      & 38.61     & 64.13            & JPL   \\
218.222188  & H$_2$CO           & 3(0,3)-2(0,2)      & 20.97     & 16.31            & CDMS  \\
218.324723  & HC$_3$N           & 24-23              &130.98     &334.19            & CDMS  \\
218.475641  & H$_2$CO           & 3(2,2)-2(2,1)      & 68.09     &  9.06            & CDMS  \\
218.732732  & c-HCCCH           & 7(1,6)-7(0,7)      & 61.17     & 33.00            & CDMS  \\
218.760066  & H$_2$CO           & 3(2,1)-2(2,0)      & 68.11     &  9.06            & CDMS  \\
218.903357  & OCS               & 18-17              & 99.81     &  9.21            & CDMS  \\
219.560328  & C$^{18}$O         & 2-1                & 15.81     &  0.02            & CDMS  \\
219.908525  & H$_2^{13}$CO      & 3(1,2)-2(1,1)      & 32.94     & 43.50            & JPL   \\
219.949442  & SO $3\Sigma$ $\nu = 0$ & 6(5)-5(4)       & 34.98     & 14.01            & CDMS  \\
220.398688  & $^{13}$CO         & 2-1                & 15.87     &  0.02            & CDMS  \\
220.709016  & CH$_3$CN $\nu = 0$& 12(3)-11(3)        &133.16     &476.93            & JPL   \\
220.730261  & CH$_3$CN $\nu = 0$& 12(2)-11(2)        & 97.44     &247.32            & JPL   \\
220.743011  & CH$_3$CN $\nu = 0$& 12(1)-11(1)        & 76.01     &252.57            & JPL   \\
220.747261  & CH$_3$CN $\nu = 0$& 12(0)-11(0)        & 68.87     &254.41            & JPL   \\
228.910489  & DNC               & 3-2                & 21.97     & 27.90           & CDMS   \\
229.347630  & SO$_2$ $\nu=0$    & 11(5,7)-12(4,8)    &122.01     &  3.13           & CDMS   \\
230.537984  & $^{12}$CO         & 2-1                & 16.57     &  0.02           & CDMS   \\
230.778011  & l-H$_2$CCC        & 11(1,10)-10(1,9)   & 79.81     &550.09           & CDMS   \\
231.060983  & OCS               & 19-18              &110.90     & 9.72            & CDMS   \\
231.220766  & $^{13}$CS         & 5-4                & 33.29     & 20.91           & CDMS   \\
231.321828  & N$_2$D$^+$ $^*$   & 3-2                & 22.20     &191.48           & CDMS   \\
231.410259  & D$_2$CO           & 4(0,4)-3(0,3)      & --        & --              & Lovas  \\
235.789641  & CO$^+$            & 2-1                & --        & --              & Lovas  \\
236.452293  & SO $3\Sigma$ $\nu = 0$ & 1(2)-2(1)     & 15.81     &  0.03           & CDMS   \\
236.512844  & HC$_3$N           & 26-25              &153.25     &362.07           & CDMS   \\
\hline
\end{tabular}
\tablefoot{An asterisk denotes the lines with hyper-fine components. Column 4: Upper state energies. Column 5: Line strengths. Column 6: The line catalogs from which the line data are taken, based on the online v3 of {\it Splatalogue} \url{http://www.cv.nrao.edu/php/splat/}.}
\end{table*}

\newpage
\begin{table*}
\begin{tabular}{lcccccc} \hline \hline
Source & date & \multicolumn{3}{c}{Calibrators} & \# antennas and & on-source telescope \\
       &      & Bandpass & Phase & Flux         & configuration  & time$^1$ (h) \\
\hline
CI\,Tau & 07-apr-2020 & 3c84 &  J0438+300 0507+179 & LkHa101 MWC349 & 10 C & 2.6 \\
        & 13-jan-2020 & 3c84 &  J0438+300 0507+179 & LkHa101 & 10 C & 4.1\\
\hline                                                 
CY\,Tau & 28-feb-2020 & 3c84 &  J0438+300 0507+179 & LkHa101 & 10 C & 2.2\\
        & 28-mar-2020 & 3c84 &  J0438+300 0507+179 & LkHa101 & 10 C & 4.5 \\
\hline                                                 
DG\,Tau & 26-mar-2020 & 3c84 &  J0438+300 0507+179 & LkHa101 & 10 C & 2.2\\
        & 06-apr-2020 & 3c84 &  J0438+300 0507+179 & LkHa101 & 10 C & 4.8 \\
\hline                                                 
DL\,Tau & 31-mar-2020 & 3c84 &  J0438+300 0507+179 & LkHa101 & 10 C & 1.9\\
        & 08-apr-2020 & 3c84 &  J0438+300 0507+179 & LkHa101 & 10 C & 4.0\\
\hline                                                 
DM\,Tau & 07-apr-2020 & 3c84 &  J0438+300 0507+179 & LkHa101 & 10 C & 3.8 \\ 
        & 09-apr-2020 & 3c84 &  J0438+300 0507+179 & LkHa101 & 10 C & 6.8 \\ 
\hline                                                 
DN\,Tau & 09-jan-2020 & 3c84 &  J0438+300 0507+179 & LkHa101 & 10 C & 3.0 \\
        & 11-jan-2020 & 3c84 &  J0438+300 0507+179 & LkHa101 & 10 C & 2.2 \\
        & 02-apr-2020 & 3c84 &  J0438+300 0507+179 & LkHa101 & 10 C & 2.0 \\
\hline                                                 
IQ\,Tau & 22-nov-2020 & 3c84 &  J0438+300 0507+179 & LkHa101 2010+723 & 10 C & 5.6 \\
\hline
UZ\,Tau\,E & 29-nov-2020 & 3c84 &  J0438+300 0507+179 & LkHa101 2010+723 & 10 C & 3.8 \\
        & 30-nov-2020 & 3C454.3 & J0438+300 0507+179 & MWC349 2010+723 & 10 C & 1.5 \\
\hline
\end{tabular}\\
$^1$ - from number of scan on source, not from uvt table after QA
\caption{Summary of the observations.}
\label{tab:summary-obs}
\end{table*}

\clearpage
\section{Channel maps, moment $0$-th maps, and Kepler plots}
\label{append:sec:obs_plots}


\begin{figure*}
\centering
\includegraphics[width=0.72\hsize,clip]{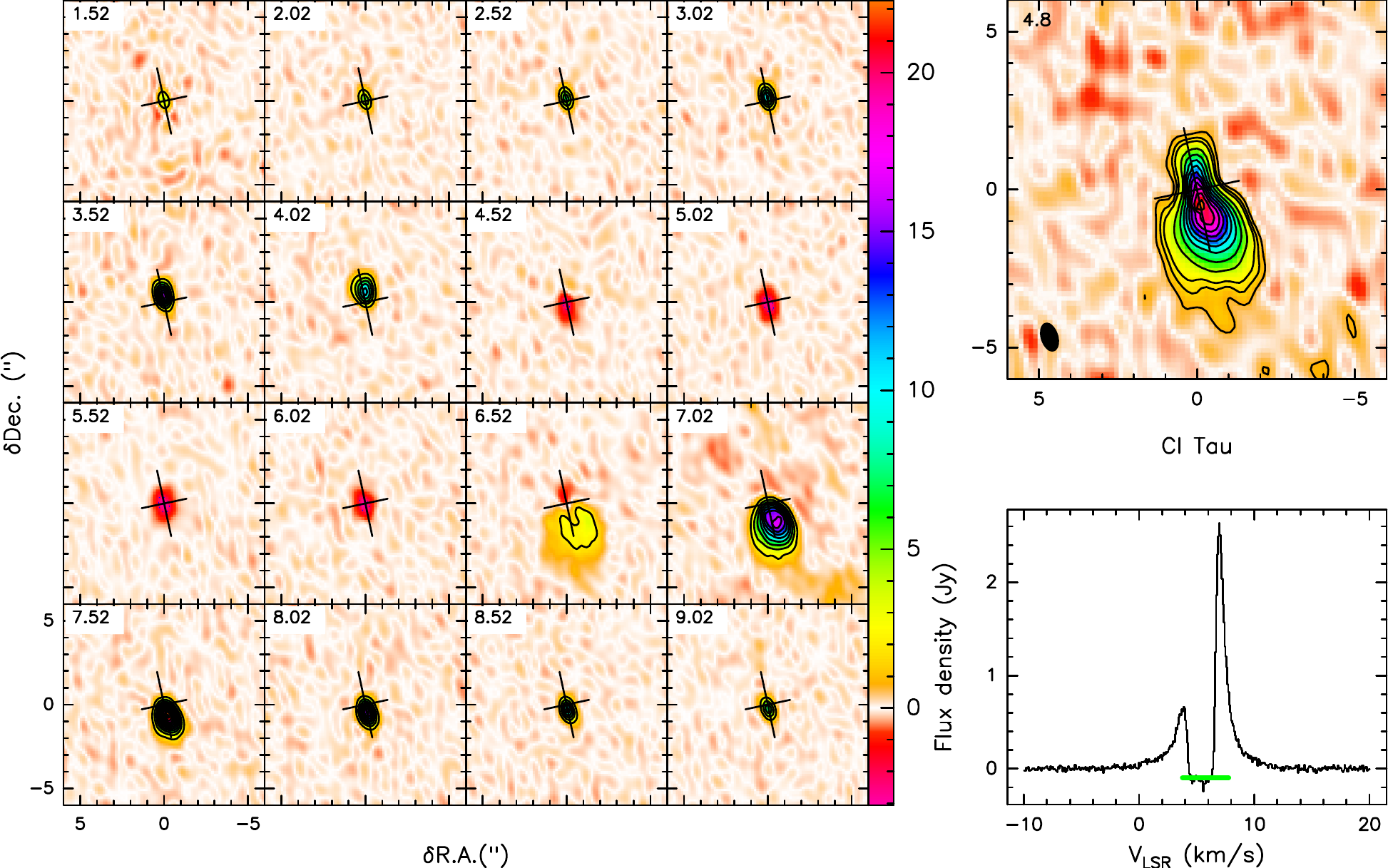}
\caption{Observed $^{12}$CO (2-1) emission in the CI~Tau disk.
Shown are the channel map of the observed line brightness distribution (top left), the moment-zero map (top right), and the integrated spectrum (top middle). The channel map shows 16 velocity channels with a step of $0.5$~km\,s$^{-1}$ in the [-4.5, +4.5]~km\,s$^{-1}$ range around the systemic velocity. The contour lines start at $2$~K, with a step of $2$~K. The color bar shows line brightness temperatures (K). The contour lines in the moment-zero plot start at 3, 6, and $9\sigma$, with a step of $6\sigma$ afterward. The $1\sigma$ rms noise (mJy\,km\,s$^{-1}$) is shown in the upper left corner of the moment-zero plot, and the synthesized beam is depicted by the dark ellipse in the left bottom corner. The $^{12}$CO~(2-1) spectrum shows a nearly absent blueshifted velocity component due to the cloud absorption.}
\label{fig:CI_Tau-CO_comp}
\end{figure*}

\begin{figure*}
\sidecaption  \includegraphics[width=12cm,clip]{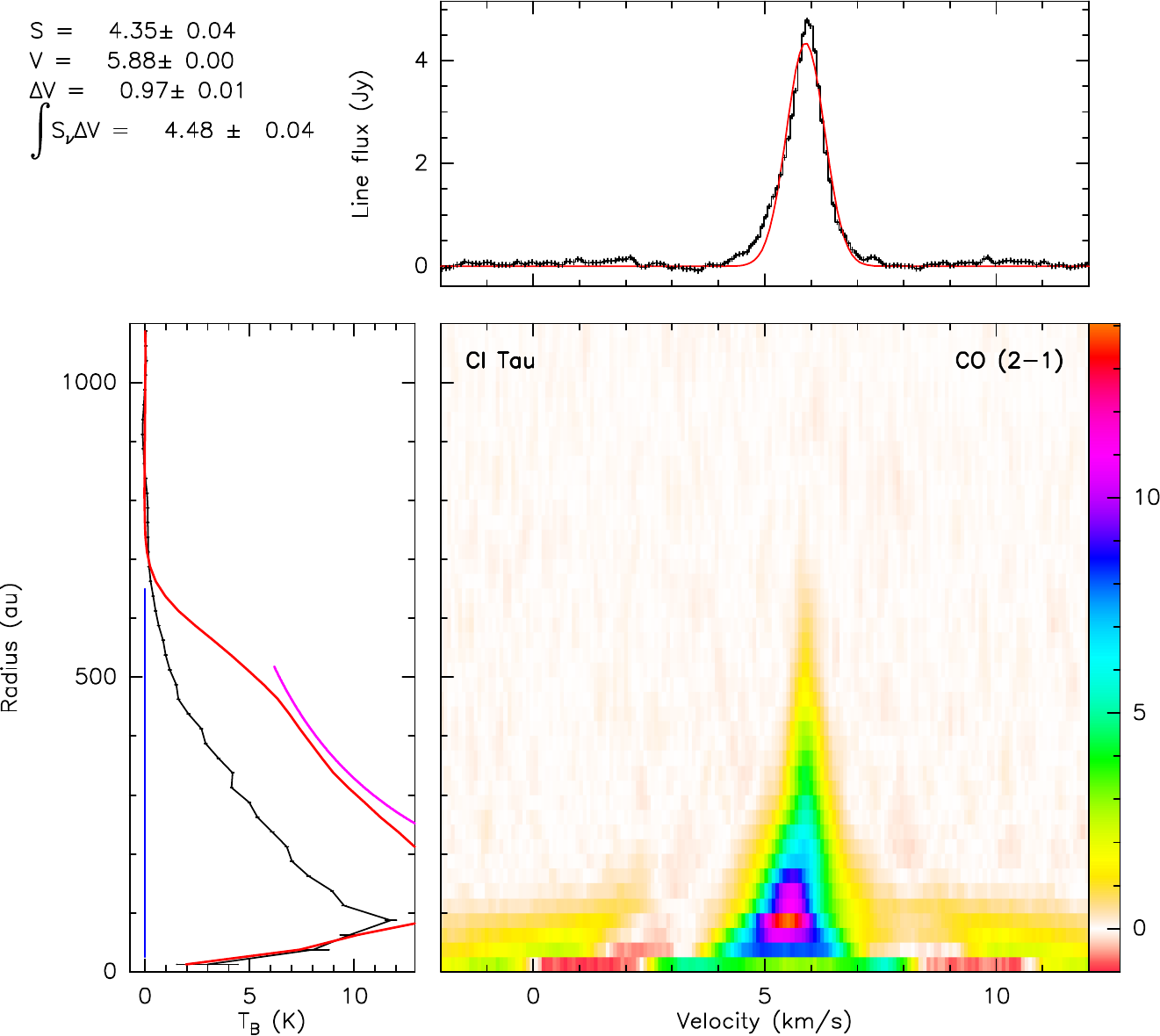}
\caption{Observations of $^{12}$CO (2-1) emission in the CI~Tau disk. Shown is a pixel-deprojected Keplerian plot consisting of the three panels: (left) the radial profile of the line brightness temperature (K), (top) the integrated spectrum (black line) overlaid with the best-fit Gaussian profile (red line), and (bottom right) aligned and stacked line intensity (K) as a function of disk radius (Y-axis; au) and velocity (X-axis; km\,s$^{-1}$). The color bar units are in Kelvin. The zigzag behavior of the $T_\mathrm{B}(r)$ curve at $\sim 200-500$~au is caused by the foreground cloud absorption.}
\label{fig:CI_Tau-CO_kepl}
\end{figure*}

\begin{figure*}
\centering
\includegraphics[width=0.72\hsize,clip]{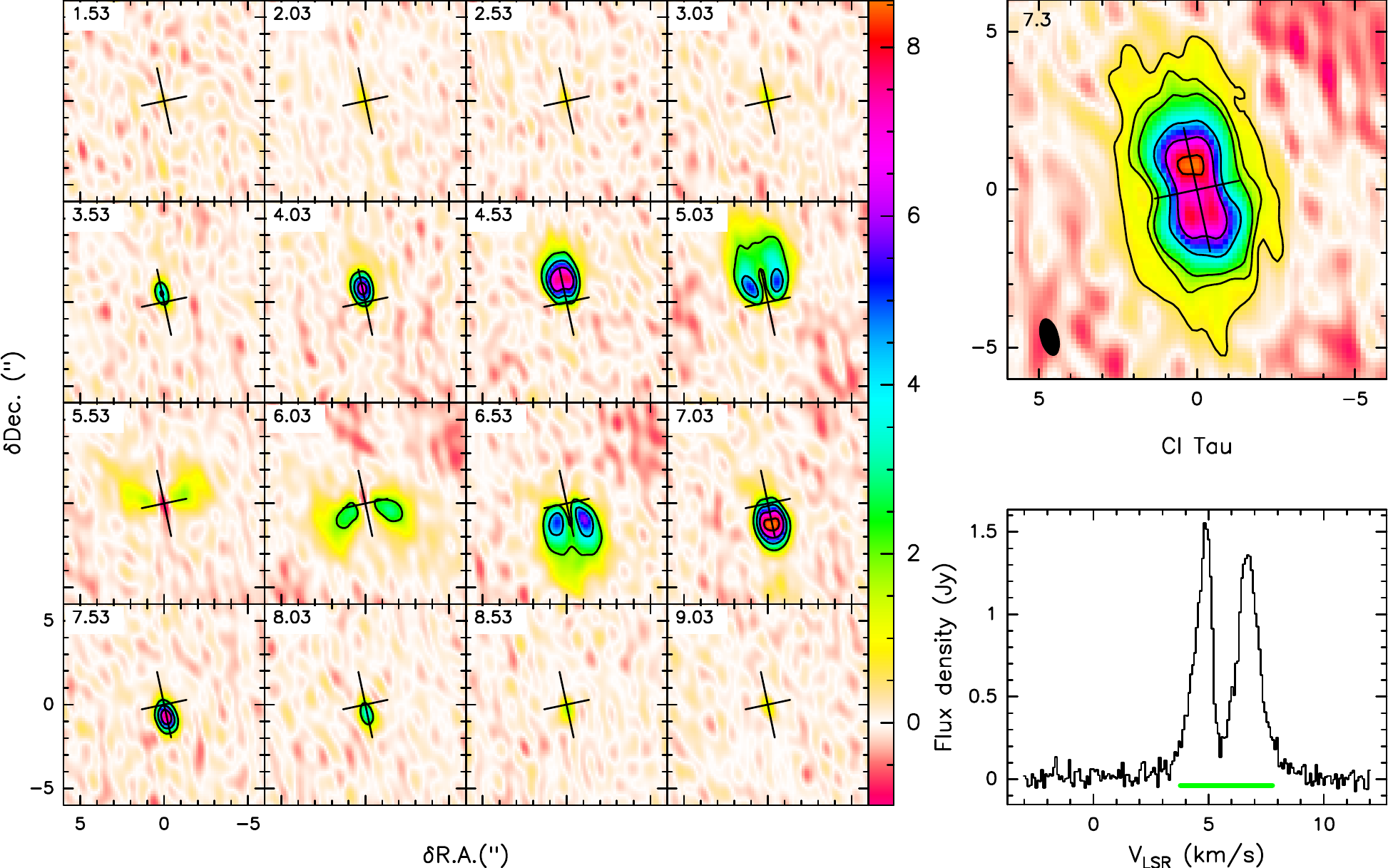}
\caption{Same as Fig.~\ref{fig:CI_Tau-CO_comp} but for $^{13}$CO (2-1) emission.}
\label{fig:CI_Tau-13CO_comp}
\end{figure*}

\begin{figure*}
\sidecaption  \includegraphics[width=12cm,clip]{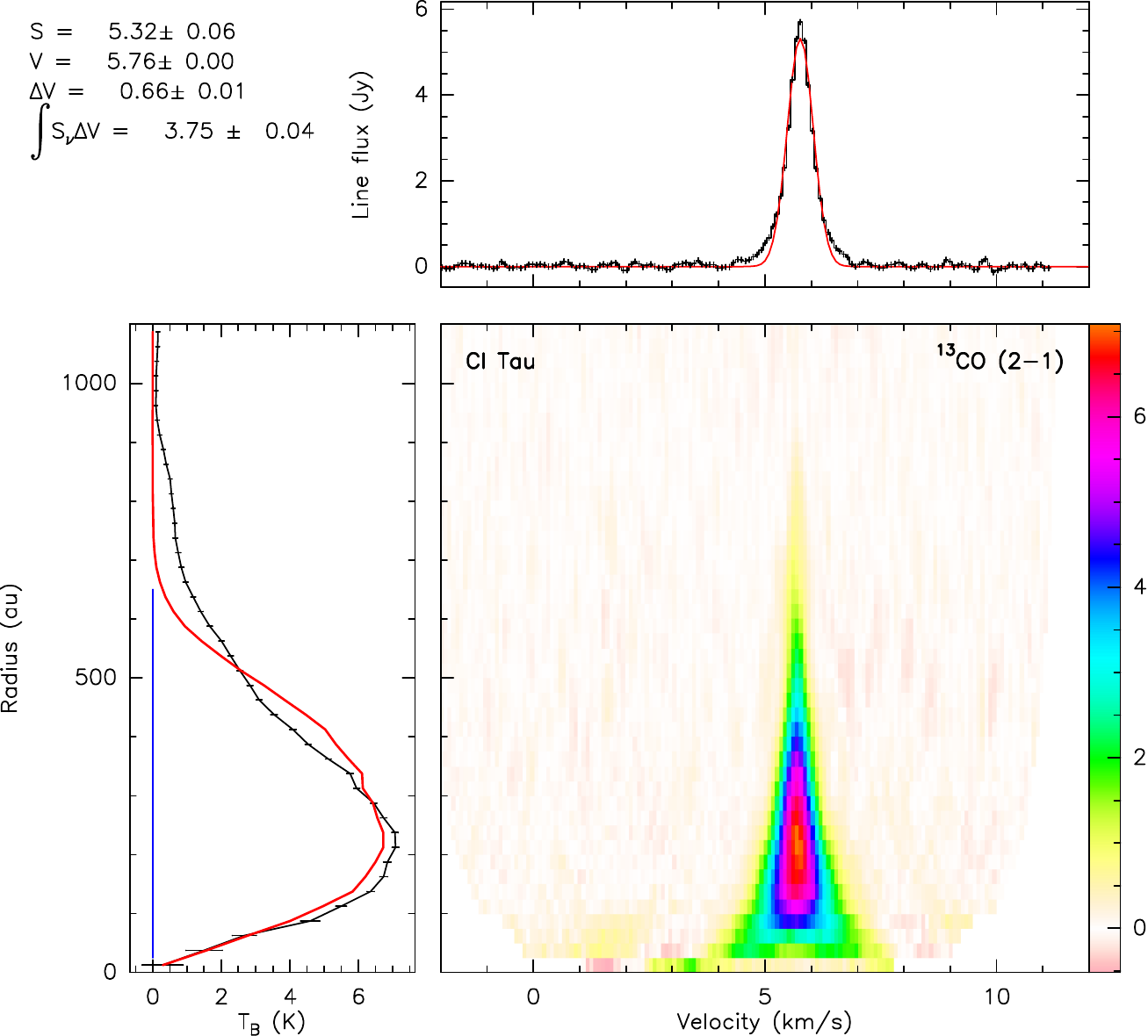}
\caption{Same as Fig.~\ref{fig:CI_Tau-CO_kepl} but for $^{13}$CO (2-1) emission.}
\label{fig:CI_Tau-13CO_kepl}
\end{figure*}

\begin{figure*}
\centering
\includegraphics[width=0.72\hsize,clip]{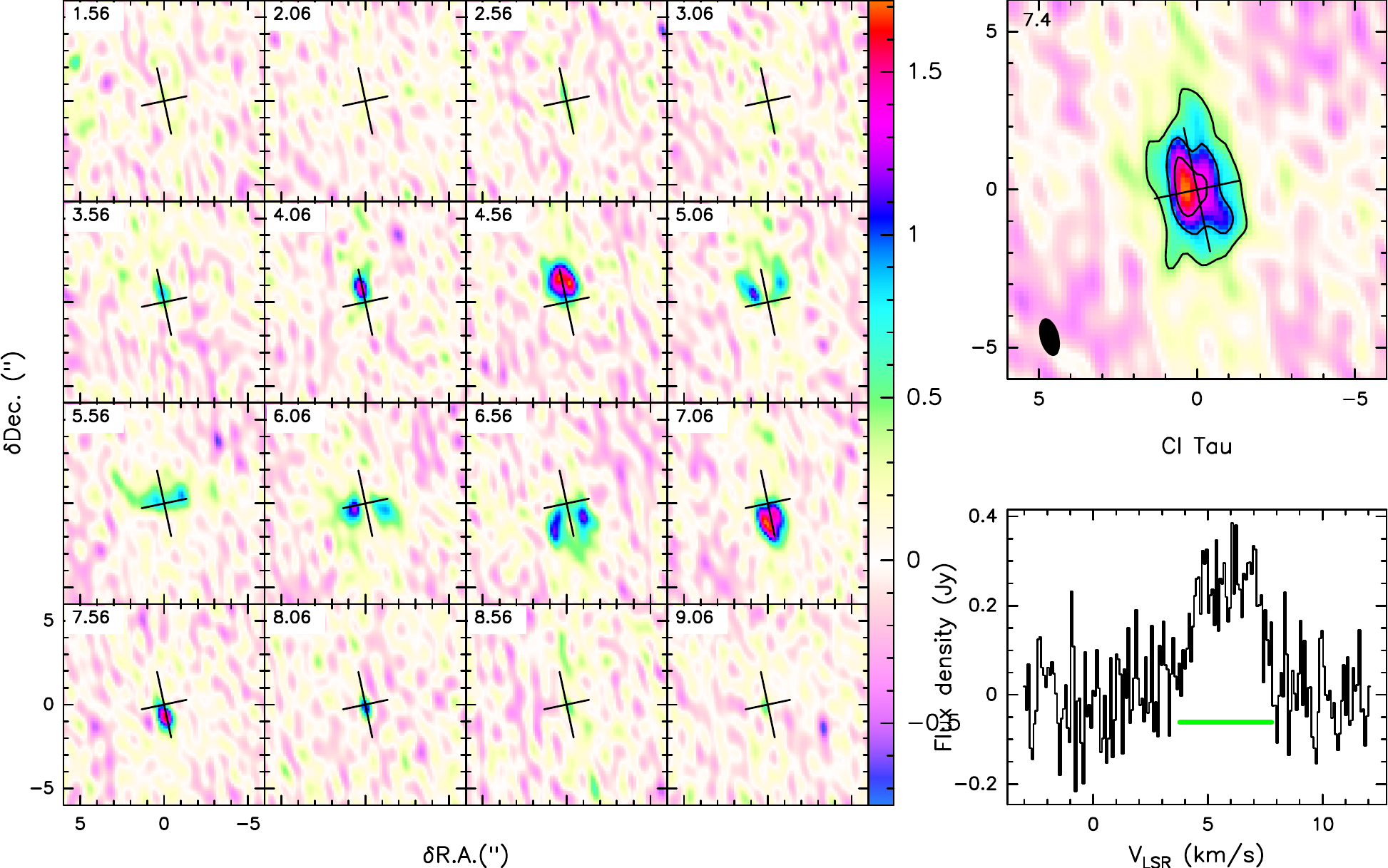}
\caption{Same as Fig.~\ref{fig:CI_Tau-CO_comp} but for C$^{18}$O (2-1) emission.}
\label{fig:CI_Tau-C18O_comp}
\end{figure*}

\begin{figure*}
\sidecaption  \includegraphics[width=12cm,clip]{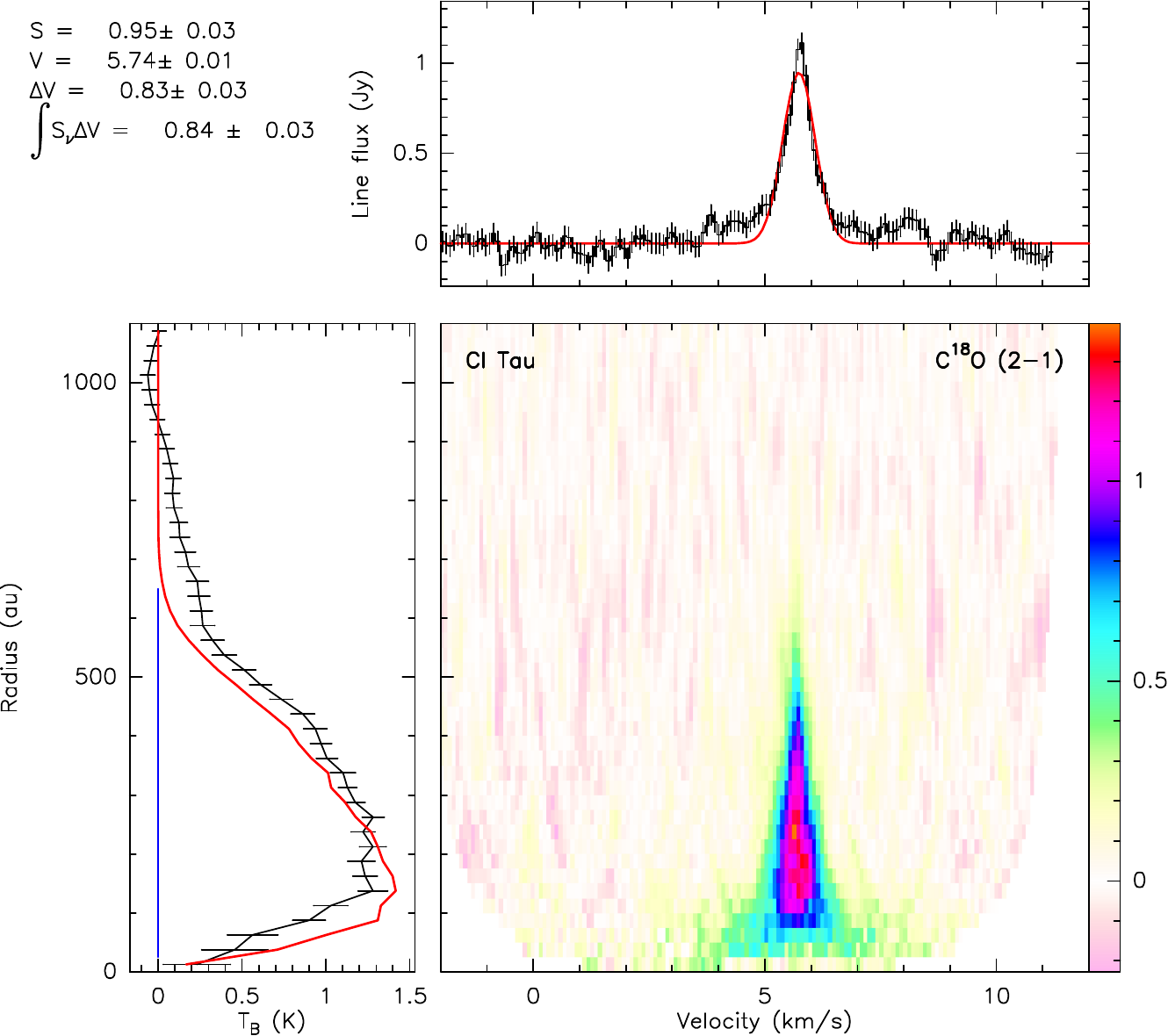}
\caption{Same as Fig.~\ref{fig:CI_Tau-CO_kepl} but for C$^{18}$O (2-1) emission.}
\label{fig:CI_Tau-C18O_kepl}
\end{figure*}

\setvalue{disk = CY_Tau}
\setvalue{name = CY~Tau}
\setvalue{mol = CO}
\setvalue{line = $^{12}$CO (2-1)}
\figcomp{\getvalue{disk}}{\getvalue{mol}}{\getvalue{line}}{\getvalue{name}}
\figkepl{\getvalue{disk}}{\getvalue{mol}}{\getvalue{line}}{\getvalue{name}}

\setvalue{mol = 13CO}
\setvalue{line = $^{13}$CO (2-1)}
\figcomp{\getvalue{disk}}{\getvalue{mol}}{\getvalue{line}}{\getvalue{name}}
\figkepl{\getvalue{disk}}{\getvalue{mol}}{\getvalue{line}}{\getvalue{name}}

\setvalue{mol = C18O}
\setvalue{line = C$^{18}$O (2-1)}
\figcomp{\getvalue{disk}}{\getvalue{mol}}{\getvalue{line}}{\getvalue{name}}
\figkepl{\getvalue{disk}}{\getvalue{mol}}{\getvalue{line}}{\getvalue{name}}

\setvalue{disk = DG_Tau}
\setvalue{name = DG~Tau}
\setvalue{mol = CO}
\setvalue{line = $^{12}$CO (2-1)}
\figcomp{\getvalue{disk}}{\getvalue{mol}}{\getvalue{line}}{\getvalue{name}}

\setvalue{mol = 13CO}
\setvalue{line = $^{13}$CO (2-1)}
\figcomp{\getvalue{disk}}{\getvalue{mol}}{\getvalue{line}}{\getvalue{name}}

\setvalue{mol = C18O}
\setvalue{line = C$^{18}$O (2-1)}
\figcomp{\getvalue{disk}}{\getvalue{mol}}{\getvalue{line}}{\getvalue{name}}

\setvalue{disk = DL_Tau}
\setvalue{name = DL~Tau}
\setvalue{mol = CO}
\setvalue{line = $^{12}$CO (2-1)}
\figcomp{\getvalue{disk}}{\getvalue{mol}}{\getvalue{line}}{\getvalue{name}}
\figkepl{\getvalue{disk}}{\getvalue{mol}}{\getvalue{line}}{\getvalue{name}}

\setvalue{mol = 13CO}
\setvalue{line = $^{13}$CO (2-1)}
\figcomp{\getvalue{disk}}{\getvalue{mol}}{\getvalue{line}}{\getvalue{name}}
\figkepl{\getvalue{disk}}{\getvalue{mol}}{\getvalue{line}}{\getvalue{name}}

\setvalue{mol = C18O}
\setvalue{line = C$^{18}$O (2-1)}
\figcomp{\getvalue{disk}}{\getvalue{mol}}{\getvalue{line}}{\getvalue{name}}
\figkepl{\getvalue{disk}}{\getvalue{mol}}{\getvalue{line}}{\getvalue{name}}

\setvalue{disk = DM_Tau}
\setvalue{name = DM~Tau}

\setvalue{mol = 13CO}
\setvalue{line = $^{13}$CO (2-1)}
\figcomp{\getvalue{disk}}{\getvalue{mol}}{\getvalue{line}}{\getvalue{name}}
\figkepl{\getvalue{disk}}{\getvalue{mol}}{\getvalue{line}}{\getvalue{name}}

\setvalue{mol = C18O}
\setvalue{line = C$^{18}$O (2-1)}
\figcomp{\getvalue{disk}}{\getvalue{mol}}{\getvalue{line}}{\getvalue{name}}
\figkepl{\getvalue{disk}}{\getvalue{mol}}{\getvalue{line}}{\getvalue{name}}

\setvalue{disk = DN_Tau}
\setvalue{name = DN~Tau}
\setvalue{mol = CO}
\setvalue{line = $^{12}$CO (2-1)}
\figcomp{\getvalue{disk}}{\getvalue{mol}}{\getvalue{line}}{\getvalue{name}}
\figkepl{\getvalue{disk}}{\getvalue{mol}}{\getvalue{line}}{\getvalue{name}}

\setvalue{mol = 13CO}
\setvalue{line = $^{13}$CO (2-1)}
\figcomp{\getvalue{disk}}{\getvalue{mol}}{\getvalue{line}}{\getvalue{name}}
\figkepl{\getvalue{disk}}{\getvalue{mol}}{\getvalue{line}}{\getvalue{name}}

\setvalue{mol = C18O}
\setvalue{line = C$^{18}$O (2-1)}
\figcomp{\getvalue{disk}}{\getvalue{mol}}{\getvalue{line}}{\getvalue{name}}
\figkepl{\getvalue{disk}}{\getvalue{mol}}{\getvalue{line}}{\getvalue{name}}

\setvalue{disk = IQ_Tau}
\setvalue{name = IQ~Tau}
\setvalue{mol = CO}
\setvalue{line = $^{12}$CO (2-1)}
\figcomp{\getvalue{disk}}{\getvalue{mol}}{\getvalue{line}}{\getvalue{name}}
\figkepl{\getvalue{disk}}{\getvalue{mol}}{\getvalue{line}}{\getvalue{name}}

\setvalue{mol = 13CO}
\setvalue{line = $^{13}$CO (2-1)}
\figcomp{\getvalue{disk}}{\getvalue{mol}}{\getvalue{line}}{\getvalue{name}}
\figkepl{\getvalue{disk}}{\getvalue{mol}}{\getvalue{line}}{\getvalue{name}}

\setvalue{mol = C18O}
\setvalue{line = C$^{18}$O (2-1)}
\figcomp{\getvalue{disk}}{\getvalue{mol}}{\getvalue{line}}{\getvalue{name}}
\figkepl{\getvalue{disk}}{\getvalue{mol}}{\getvalue{line}}{\getvalue{name}}

\setvalue{disk = UZ_Tau}
\setvalue{name = UZ~Tau~E}
\setvalue{mol = CO}
\setvalue{line = $^{12}$CO (2-1)}
\figcomp{\getvalue{disk}}{\getvalue{mol}}{\getvalue{line}}{\getvalue{name}}
\figkepl{\getvalue{disk}}{\getvalue{mol}}{\getvalue{line}}{\getvalue{name}}

\setvalue{mol = 13CO}
\setvalue{line = $^{13}$CO (2-1)}
\figcomp{\getvalue{disk}}{\getvalue{mol}}{\getvalue{line}}{\getvalue{name}}
\figkepl{\getvalue{disk}}{\getvalue{mol}}{\getvalue{line}}{\getvalue{name}}

\setvalue{mol = C18O}
\setvalue{line = C$^{18}$O (2-1)}
\figcomp{\getvalue{disk}}{\getvalue{mol}}{\getvalue{line}}{\getvalue{name}}
\figkepl{\getvalue{disk}}{\getvalue{mol}}{\getvalue{line}}{\getvalue{name}}

\end{appendix}


\end{document}